\documentclass[aps,prd,preprint,a4paper,showpacs,nofootinbib,superscriptaddress]{revtex4-2}
\usepackage{bm}
\usepackage{indentfirst}
\usepackage{amsmath}
\usepackage{graphicx}
\usepackage{float}
\usepackage{amssymb}
\usepackage{subfigure}
\usepackage{amssymb}
\usepackage{hyperref}
\usepackage{epstopdf}
\usepackage[section]{placeins}

\usepackage[utf8]{inputenc}
\hypersetup{
    colorlinks=true,
    linkcolor=red,
    citecolor=blue,
}
\usepackage{color}
\usepackage[T1]{fontenc}
\usepackage{txfonts}
\usepackage{orcidlink}

\usepackage{adjustbox,lipsum}

\newcommand{\Rmnum}[1]{\expandafter\@slowromancap\romannumeral #1@} 
\newcommand{\bq}{\begin{equation}}
\newcommand{\eq}{\end{equation}}
\newcommand{\bqn}{\begin{eqnarray}}
\newcommand{\eqn}{\end{eqnarray}}
\newcommand{\nb}{\nonumber}

\DeclareMathOperator{\sech}{sech}

\makeatother

\begin{document}

\title{On the mapping between bound states and black hole quasinormal modes via analytic continuation: a spectral instability perspective}

\author{Guan-Ru Li}
\affiliation{Faculdade de Engenharia de Guaratinguet\'a, Universidade Estadual Paulista, 12516-410, Guaratinguet\'a, SP, Brazil}

\author{Wei-Liang Qian}\email[E-mail: ]{wlqian@usp.br (corresponding author)}
\affiliation{Escola de Engenharia de Lorena, Universidade de S\~ao Paulo, 12602-810, Lorena, SP, Brazil}
\affiliation{Faculdade de Engenharia de Guaratinguet\'a, Universidade Estadual Paulista, 12516-410, Guaratinguet\'a, SP, Brazil}
\affiliation{Center for Gravitation and Cosmology, School of Physical Science and Technology, Yangzhou University, Yangzhou 225009, China}

\author{Xiao-Mei Kuang}
\affiliation{Center for Gravitation and Cosmology, School of Physical Science and Technology, Yangzhou University, Yangzhou 225009, China}

\author{Ramin G. Daghigh}
\affiliation{Natural Sciences Department, Metropolitan State University, Saint Paul, Minnesota, 55106, USA}

\author{Jodin C. Morey}
\affiliation{Le Moyne College, Syracuse, New York, 13214, USA}

\author{Michael D. Green}
\affiliation{Mathematics and Statistics Department, Metropolitan State University, Saint Paul, Minnesota, 55106, USA}

\author{Peng Xu}
\affiliation{Center for Gravitational Wave Experiment, National Microgravity Laboratory, Institute of Mechanics, Chinese Academy of Sciences, Beijing 100190, China}

\author{Rui-Hong Yue}
\affiliation{Center for Gravitation and Cosmology, School of Physical Science and Technology, Yangzhou University, Yangzhou 225009, China}

\begin{abstract}
In this work, we investigate the relation between bound states and quasinormal modes within black hole perturbation theory in the context of spectral instability.
Our analysis indicates that the reliability of such spectral mapping extends beyond the domain of validity of the analytic continuation employed to connect the perturbative bound-state problem to the corresponding open-system dynamics.
Moreover, we demonstrate that, for the numerical scheme proposed by V\"olkel to work, the transformations of the metric parameters must be carried out within the radius of convergence of the underlying Taylor expansion.
As analytically accessible explicit examples, we explore the perturbed delta-function and P\"oschl-Teller potential barriers. 
For the latter, we construct two distinct perturbative setups for which the convergence of the series expansion involved in the perturbation theory can be rigorously controlled.
When the deformation is placed near the potential's extremum, the resulting corrections to the bound-state energies can be analytically continued to yield perturbed quasinormal frequencies, in agreement with known semi-analytic results.
In contrast, when the perturbation is localized asymptotically far from the compact object, the bound states are only mildly modified and are accurately described by a perturbative expansion to the first order. 
However, the associated analytic continuation yields a strongly deformed spectrum that shows no clear connection to the quasinormal modes.
These findings contribute to the effort to scrutinize the conditions under which bound states faithfully encode quasinormal spectra and to shed light on the underlying physics of black hole spectral instability.
\end{abstract}

\date{Oct. 13th, 2026}

\maketitle

\newpage
\section{Introduction}\label{section1}

Quasinormal modes (QNMs) are the characteristic damped oscillations that govern the relaxation of perturbed black holes and compact objects~\cite{agr-qnm-review-01, agr-qnm-review-02, agr-qnm-review-03, agr-qnm-review-04}, and they provide a crucial framework for describing the late-time, ringdown phase of binary black hole coalescences~\cite{agr-BH-spectroscopy-005, agr-BH-spectroscopy-006, agr-BH-spectroscopy-015, agr-BH-spectroscopy-016, agr-BH-spectroscopy-035, agr-BH-spectroscopy-038, agr-BH-spectroscopy-042, agr-BH-spectroscopy-061}.
With the advent of gravitational-wave astronomy and the detection of black hole mergers by ground-based~\cite{agr-LIGO-01, agr-LIGO-02} and future space-borne interferometers~\cite{agr-LISA-01, agr-TianQin-01, agr-Taiji-01}, the accurate modeling of QNMs has become a cornerstone of tests of general relativity and black hole spectroscopy~\cite{agr-BH-spectroscopy-review-04}.
In particular, high overtones have attracted considerable interest, in part because of Hod's conjecture~\cite{agr-qnm-022}, which was built on earlier numerical results by Nollert~\cite{agr-qnm-continued-fraction-12} for the highly damped QNMs of Schwarzschild black holes.  This conjecture was refined later by Maggiore~\cite{agr-qnm-023},  which relates the asymptotic spacing between consecutive modes to the quantization of black hole area and entropy via Bohr's correspondence principle.  

More recently, following the works of Nollert and Price~\cite{agr-qnm-instability-002, agr-qnm-instability-003} and of Aguirregabiria and Vishveshwara~\cite{agr-qnm-027, agr-qnm-030}, it has been shown~\cite{agr-qnm-instability-011, agr-qnm-Poschl-Teller-16, agr-qnm-echoes-20} that even tiny metric perturbations far from the compact object can dramatically affect the higher overtones.  
Beyond invalidating Hod's original conjecture, this result challenges the conventional wisdom that small deformations of the effective potential induce only minor changes in the QNM spectrum.  
The resulting high-overtone modes are found to migrate toward the real frequency axis, in sharp contrast with the typical shift toward larger imaginary parts observed in most black hole backgrounds~\cite{agr-qnm-continued-fraction-12, agr-qnm-continued-fraction-23}.  

From a broader perspective on spectral instability, Jaramillo {\it et al.}~\cite{agr-qnm-instability-007, agr-qnm-instability-013} investigated these effects by analyzing random perturbations of the effective potential using pseudospectrum techniques in black hole perturbation theory.  
Their results point to a universal instability of high-overtone modes triggered by ultraviolet, i.e.~small-scale, perturbations.  
Notably, Cheung {\it et al.}~\cite{agr-qnm-instability-015} showed that even the fundamental mode can become unstable under smooth localized perturbations, including Gaussian bumps.
Such bumps introduce no discontinuity and have rapidly decaying spatial Fourier components, suppressing the ultraviolet content associated with sharp discontinuities.
This suppression can moderate spectral instability, although smoothness alone does not guarantee spectral stability (see also Ref.~\cite{spectral-instability-review-20}).
In this context, spectral instability and its consequences may have important implications for observational astrophysics, particularly for black hole spectroscopy~\cite{agr-BH-spectroscopy-review-04}, since realistic gravitational-wave sources are rarely isolated and typically interact with surrounding matter and fields.

Among various approaches for evaluating QNMs, a method originally proposed by Blome, Ferrari, and Mashhoon~\cite{agr-qnm-Poschl-Teller-01, agr-qnm-Poschl-Teller-02, agr-qnm-Poschl-Teller-03} is particularly intriguing.  
It was proposed that, when an analytic continuation of the relevant parameters is feasible, the quasinormal frequencies of a black hole can be related to the bound-state spectrum of an associated problem, in which a potential barrier is effectively mapped into a potential well.  
Besides being one of the very few analytic strategies available for tackling the QNM problem, this framework also sheds light on the generic relation between resonance states and bound states through the WKB approximation, as pointed out by Hatsuda~\cite{agr-qnm-WKB-20, agr-qnm-WKB-22, agr-qnm-WKB-23}.  
Nonetheless, due to its purely analytic nature, the application of the method is limited to a small class of potentials whose bound-state spectra are known in closed-analytic form.  
In this regard, V\"olkel~\cite{agr-qnm-Poschl-Teller-20} proposed to circumvent this restriction by introducing a numerical scheme that relies only on the values of the effective potential and its derivatives with respect to the relevant parameters.  
Since computing bound states numerically is often easier than directly calculating the corresponding QNMs, this approach is also appealing from a practical standpoint.  
The proposed method is based on a Taylor expansion of the bound-state energies in terms of the relevant parameters, followed by analytic continuation, and it was argued~\cite{agr-qnm-Poschl-Teller-20} that the mapped parameters need not remain close to their original values for the expansion to remain valid.  
This claim was illustrated for the original P\"oschl-Teller potential, where numerical results could be benchmarked against exact expressions, and convergence was manifestly obtained.  
Furthermore, it was suggested that this strategy could be employed to investigate spectral instability.  
In particular, Refs.~\cite{agr-qnm-Poschl-Teller-21, agr-qnm-Poschl-Teller-23} analyzed the bound-state spectrum and wavefunctions of the inverted Regge-Wheeler potential, and compared them with those of a slightly deformed potential well, in a setup reminiscent of studies on the stability of the fundamental mode~\cite{agr-qnm-instability-015, agr-qnm-instability-058, agr-qnm-instability-056, agr-qnm-instability-055}.  
However, perhaps due to the numerical challenges associated with computing higher-order derivatives, the corresponding QNMs were not obtained.  
The inability to find the modes, together with open questions about the validity of the analytic continuation itself, has led to some ambiguity and subsequent criticism~\cite{agr-qnm-Poschl-Teller-19, agr-qnm-Poschl-Teller-22}.  
It is worth noting that, for small deformations of the effective potential, perturbative schemes tailored to non-Hermitian systems have been developed to compute the resulting QNMs, either via generalized logarithmic perturbation theory~\cite{agr-qnm-033, agr-qnm-034} or by combining the inverted potential with analytic continuation~\cite{agr-qnm-Poschl-Teller-18}.  
Existing applications, however, have predominantly focused on regimes where the deformation of the QNMs in the complex-frequency plane remains modest.  
In contrast, the regime of spectral instability typically involves sizable displacements of QNMs and might push the perturbative methods beyond their domain of validity, so that the feasibility of analytic continuation in this context is not entirely settled.

The present study is motivated by the above considerations.  
We aim to assess the viability of the mapping between bound states and QNMs in the context of spectral instability.  
Specifically, it is instructive to compare analytic expressions for the quasinormal frequencies obtained via analytic continuation of the bound-state spectrum with those computed directly using black hole perturbation theory.  
To this end, we resort to two simplified setups in which the problem admits an analytic treatment.
Specifically, we explored the perturbed QNM spectrum of the delta-function potential barrier~\cite{agr-qnm-instability-065, agr-qnm-echoes-50} and the modified P\"oschl-Teller effective potential previously explored in~\cite{agr-qnm-Poschl-Teller-06, agr-qnm-Poschl-Teller-07, agr-qnm-Poschl-Teller-16, agr-qnm-Poschl-Teller-17}.
For the latter, we consider two scenarios for which the bound-state problem is fully under control, in the sense that the matrix element of the first-order energy correction is manifestly convergent.  
This allows us to apply complex analytic continuation to derive the corresponding QNMs and to compare the results with existing calculations.

The remainder of the paper is organized as follows.  
After briefly revisiting the analytic continuation scheme for QNMs in Sec.~\ref{section2}, and in particular, its proposed implementation via numerical approach, then
in Sec.~\ref{section3} we apply the scheme to the simplest possible scenario of a perturbed delta-function potential barrier.
Subsequently, the analysis is carried out for the modified P\"oschl-Teller effective potential. 
We analyze the bound states of the modified P\"oschl-Teller effective potential in Sec.~\ref{section3}, where two scenarios are discussed in detail.  
Subsequently, we perform the analytic continuation to obtain the QNMs in Sec.~\ref{section4}, and in Sec.~\ref{section5} we compare these results with those available in the literature.  
Further discussion and concluding remarks are presented in the final section.

\section{Mapping between bound states and QNMs}\label{section2}

In this section, we briefly review the scheme of analytic continuation for evaluating QNMs using the solutions of the corresponding bound-state problem.
By focusing on background spacetimes subject to specific symmetry, the study of black hole perturbation theory often leads to the following master equation~\cite{agr-qnm-review-01, agr-qnm-review-02},
\begin{eqnarray}
\frac{\partial^2}{\partial t^2}\Phi(t, x)+\left(-\frac{\partial^2}{\partial x^2}+V_\mathrm{eff}\right)\Phi(t, x)=0 ,
\label{master_eq_ns}
\end{eqnarray}
where the spatial coordinate $x\in(-\infty,+\infty)$ is known as the tortoise coordinate, and the effective potential $V_\mathrm{eff}$ is governed by the given spacetime metric.  
However, $V_\mathrm{eff}$ is an effective potential that typically depends on the spin $\bar{s}$, and angular momentum $\ell$, of the wavefunction.
For instance, the Regge-Wheeler potential $V_\mathrm{RW}$ for the Schwarzschild black hole metric is
\bqn
V_\mathrm{RW}=f(r)\left[\frac{\ell(\ell+1)}{r^2}+(1-{\bar{s}}^2)\frac{r_h}{r^3}\right],
\label{Veff_RW}
\eqn
where 
\bqn
f(r)=1-r_h/r ,
\label{f_RW}
\eqn
and $r_h=2M$ is the event horizon radius determined by the black hole mass $M$.  
The tortoise coordinate is related to the original radial coordinate $r\in (r_h, +\infty)$ by the relation 
\bqn
x=\int \frac{dr}{f(r)} = r+r_h \ln\left(\frac{r}{r_h}-1\right) .
\eqn
Here, we use geometric units with $G=c=1$.
The P\"ochl-Teller effective potential is given by the form
\bqn
V_\mathrm{PT}(x)={V_0}~{\sech}^2\left(\frac{x}{b}\right),
\label{Veff_PT}
\eqn
where $V_0$ and $b$ are two parameters governing the shape of the potential.
It is noted that both Eqs.~\eqref{Veff_RW} and~\eqref{Veff_PT} possess the form of a potential barrier.

By Fourier transforming to the frequency domain,
\bqn
\Phi(t,x)=(2\pi)^{-1}\int d\omega \, e^{-i\omega t}\Psi(\omega,x) ,\nb
\eqn
Eq.~\eqref{master_eq_ns} becomes
\begin{equation}
\frac{d^2\Psi(\omega, x)}{dx^2}
+\bigl[\omega^2-V_\mathrm{eff}\bigr]\Psi(\omega, x) = 0 , 
\label{master_frequency_domain}
\end{equation}
subject to the boundary conditions
\begin{equation}
\Psi(\omega, x) \sim
\begin{cases}
   e^{-i\omega x}, &  x \to -\infty, \\
   e^{+i\omega x}, &  x \to +\infty,
\end{cases}
\label{master_bc0}
\end{equation}
for which $\omega$ takes discrete complex values $\omega_n$, known as the quasinormal frequencies, with $n$ denoting the overtone number.  
Because of the boundary conditions in Eq.~\eqref{master_bc0}, and the fact that $V_\mathrm{eff}$ represents a potential barrier, QNMs can naturally be interpreted as resonance states in a one-dimensional scattering problem.

The method introduced by Blome, Ferrari, and Mashhoon~\cite{agr-qnm-Poschl-Teller-01, agr-qnm-Poschl-Teller-02, agr-qnm-Poschl-Teller-03} proposes to solve, instead of Eq.~\eqref{master_frequency_domain}, the Schr\"odinger equation (with $\hbar^2/2m=1$),
\begin{equation}\label{bound_frequency_domain} 
\left(-\frac{d^2}{dx^2}+\widetilde{V}_\mathrm{eff}\right)\widetilde{\Psi}(\omega, x) = E\,\widetilde{\Psi}(\omega, x) , 
\end{equation}
with real eigenvalues $E_n$, subject to the bound-state boundary conditions
\begin{equation}
\widetilde{\Psi}(\omega, x) \sim
\begin{cases}
   e^{+\omega x}, &  x \to -\infty , \\
   e^{-\omega x}, &  x \to +\infty ,
\end{cases}
\label{master_bc1}
\end{equation}
where $\omega=\sqrt{-E}>0$ for $E<\widetilde{V}_\mathrm{eff}(\pm\infty) = 0$.

It is observed that the solutions of Eq.~\eqref{master_frequency_domain} can be formally related to those of Eq.~\eqref{bound_frequency_domain} by the following set of transformations.  
First, boundary conditions Eqs.~\eqref{master_bc0} can be mapped to~\eqref{master_bc1} if the spatial coordinate $x$ is subjected to a Wick rotation
\bqn
x \to i x .\label{WickRot}
\eqn
Second, one must effectively flip the barrier $V_\mathrm{eff}$ upside down, i.e.~$\widetilde{V}_\text{eff}=-V_\text{eff}$, into a potential well so that the Schrödinger equation~\eqref{bound_frequency_domain} defines a genuine bound-state problem.  
This is achieved if the effective potential ${\widetilde V}_\mathrm{eff}$ is obtained by transforming the parameters in Eq.~\eqref{master_frequency_domain} according to
\bqn
P_i \to \pi\left(P_i\right),\label{transP}
\eqn
such that the transformed effective potential satisfies
\bqn \label{Vcondi}
V_\mathrm{eff}(ix,  \{\pi(P_i)\})
= V_\mathrm{eff}(x, \{P_i\}) .
\eqn
After solving for the bound-state energies $E_n$ of the potential well $\widetilde{V}_\mathrm{eff}$, and assuming that $E_n$ is analytic in the parameters, one may apply the inverse transformations of Eq.~\eqref{transP},
to the arguments of $E_n(\{P_i\})$, and finally define
\bqn
\omega_n = \pm \sqrt{-E_n\left(\left\{\pi^{-1}\left(P_i\right)\right\}\right)} ,\label{MapQNM}
\eqn
where the sign in front of the square root is chosen to ensure $\mathrm{Im}\,\omega_n < 0$.

Since Eqs.~\eqref{transP} and~\eqref{MapQNM} imply an analytic assessment of the bound-state problem, it has been understood that the applicability of this method is restricted to cases in which a closed-form analytic expression for the bound-state energy spectrum is available.
In~\cite{agr-qnm-Poschl-Teller-20}, V\"olkel proposed to circumvent this restriction by introducing a numerical scheme that relies only on the values of the effective potential and its derivatives with respect to the relevant parameters.
Specifically, one views $E_n$ as a multivariable real function $E_n(\{Q_i\})$ whose independent variables $\{Q_i\}$ take (real) values $Q_{0,i}$, and Taylor expands it around the latter
\begin{equation}\label{TaylorEn}
E_n(\{Q_i\})
=
E_n(\{Q_{0,i}\})
+
\sum_{j}
\left.
\frac{\partial E_n}{\partial Q_j}
\right|_{\{Q_{0,i}\}}
(Q_j-Q_{0,j})
+
\frac{1}{2}
\sum_{j}
\sum_{k}
\left.
\frac{\partial^2 E_n}{\partial Q_j \partial Q_k}
\right|_{\{Q_{0,i}\}}
(Q_j-Q_{0,j})(Q_k-Q_{0,k})
+
\cdots .
\end{equation}
Assuming that $E_n(\{Q_i\})$ is an analytic function in the vicinity of $\{Q_{0,i}\}$, and since $\{Q_{0,i}\}$ is not a pole, the Laurent expansion reduces to Eq.~\eqref{TaylorEn}, i.e., a Taylor expansion in the complex plane. 
In particular, the complex partial derivatives in the expansion coincide with the (real) partial derivatives, which can be evaluated numerically.
Now, if $\left\{\pi^{-1}\left(Q_{0,i}\right)\right\}$ lies within the radius of convergence, we can substitute $Q_i=\pi^{-1}(Q_{0,i})$ into Eq.~\eqref{TaylorEn} and expect to find the corresponding QNMs.
Since computing bound states numerically is often easier than directly calculating the corresponding QNMs, this approach is appealing from a practical standpoint.

For the Regge-Wheeler effective potential Eq.~\eqref{Veff_RW}, one introduces an overall factor $\kappa$~\cite{agr-qnm-Poschl-Teller-02, agr-qnm-Poschl-Teller-03, agr-qnm-Poschl-Teller-20},
\bqn
V^{(\kappa)}_\mathrm{RW} \equiv \kappa V_\mathrm{RW} ,
\label{Veff_RW_lambda}
\eqn
and the following transformations satisfy the requirement Eq.~\eqref{Vcondi}:
\begin{eqnarray}
\pi(\kappa) &=& -\kappa, \nb\\
\pi(r_h) &=& i r_h, \nb\\
\pi(\ell) &=& \ell, \nb\\
\pi(\bar{s}) &=& \bar{s} .\label{piRW}
\end{eqnarray}
In this case, Eq.~\eqref{WickRot} is implied by $r\to ir$.
Effectively, the potential $V_\mathrm{RW}$ changes sign and all length-like quantities are multiplied by a factor $i$.

The simplest example is the P\"oschl-Teller effective potential Eq.~\eqref{Veff_PT}.  
In addition to Eq.~\eqref{WickRot}, one introduces the following transform
\begin{eqnarray}
\pi(V_0) &=& V_0, \nb\\
\pi(b) &=& i b .\label{piPT}
\end{eqnarray}

The bound-state energies of the P\"oschl-Teller potential well are known analytically~\cite{book-quantum-mechanics-Flugge},
\bqn
E_n (V_0, b) = -\Omega_{n}^2 ,\label{EnPT}
\eqn
with
\bqn
\Omega_{n}=\frac{1}{b}\left(\lambda-n\right),\label{OmegaPT}
\eqn
and
\bqn
\lambda=-\frac{1}{2}+\frac{1}{2}\sqrt{1+4b^2V_0} .\label{lambdaPT}
\eqn

Eq.~\eqref{piPT} implies
\bqn
\lambda\to \lambda\left(\pi^{-1}(b)\right)=-\frac{1}{2}+\frac{1}{2}\sqrt{1-4b^2V_0} .\label{piLambdaPT}
\eqn
Substituting this and Eqs.~\eqref{EnPT} and~\eqref{OmegaPT} into Eq.~\eqref{MapQNM}, and choosing the plus sign, one recovers the well-known QNM spectrum
\bqn
\omega_n 
= \Omega_n\left(\pi^{-1}(V_0), \pi^{-1}(b)\right)
= \sqrt{V_0-\frac{1}{4b^2}}-i\,\frac{1}{b}\left(n+\frac{1}{2}\right) , 
\label{QNM_PT}
\eqn
where we note that the QNMs have a nonvanishing real part as long as
\bqn
V_{0} > \frac{1}{4b^2} .\label{condPT}
\eqn
Otherwise, the QNMs are purely imaginary
\bqn
\omega_n = \frac{i}{b}\left(-\frac{1}{2}+\frac{1}{2}\sqrt{1-4b^2V_0}-n\right) . 
\label{QNM_PT2}
\eqn

Regarding the expansion Eq.~\eqref{TaylorEn}, one can evaluate the radius of convergence in the complex plane of $b$.
In fact, the analytic function $\Omega_n(b)\sim\sqrt{1/b^2+4V_0}$ has a brach cut between $b_\pm=\pm i/\sqrt{4V_0}$ and a pole at $b_0=0$.
Therefore, the radius of convergence about a point $b$ on the positive real axis is 
\bqn
\mathcal{R}(b)=\mathrm{min}\left(\sqrt{b^2+\frac{1}{4V_0}}, b\right)=b .
\eqn
Therefore, the mapping $b \to \pi^{-1}(b) = - i b$ induces a displacement $\sqrt{2}\,b > \mathcal{R}(b)$, which clearly lies outside the radius of convergence when $\Omega_n(b)$ is regarded as an analytic function expanded around a real value of $b$. 
It is rather intriguing that, despite this, the analytic prescription remains valid: a ``reckless'' application of the analytic continuation $b \to \pi^{-1}(b)$ correctly reproduces the QNM spectrum.

The above dilemma is somewhat alleviated by considering the parameter $\alpha = 1/b$ instead.
The corresponding analytic function $\Omega_n(\alpha)\sim\sqrt{\alpha^2+4V_0}$ has a branch cut between $\alpha_\pm=\pm i\sqrt{4V_0}$.
Subsequently, the radius of convergence about a point $\alpha$ on the positive real axis is
\bqn
\mathcal{R}(\alpha)=\sqrt{\alpha^2+4V_0} .\label{eq:Ralpha}
\eqn
Given the condition Eq.~\eqref{condPT}, namely $V_0>\alpha^2/4$, the mapping $\alpha \rightarrow \pi^{-1}(\alpha) = i\alpha$ manifestly lies inside the region of convergence.
This is precisely the circumstance under which the numerical scheme proposed in Ref.~\cite{agr-qnm-Poschl-Teller-20} was carried out.

\begin{figure}[htp]
\centering
\includegraphics[scale=0.49]{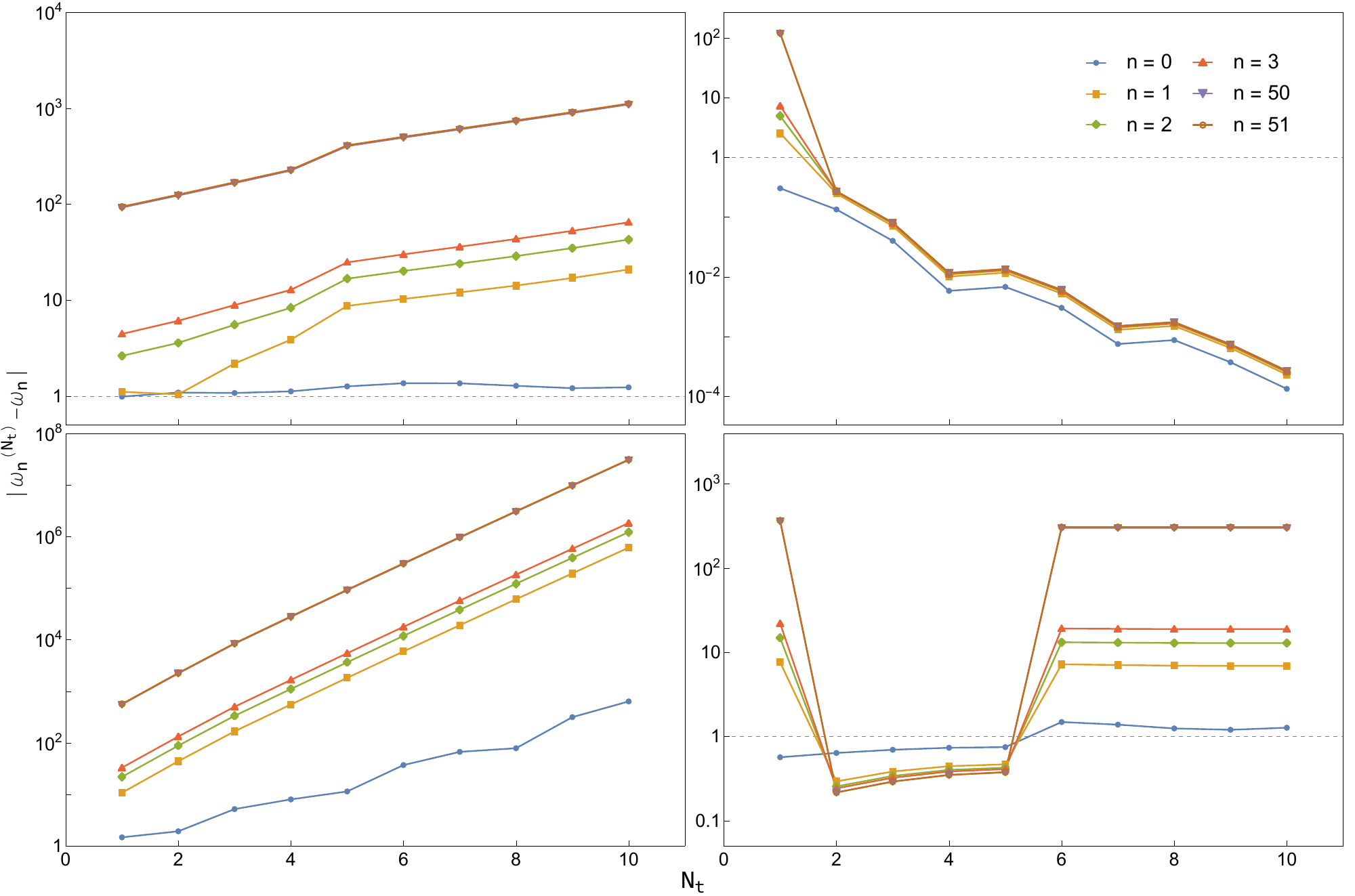}
\caption{
The deviations of the calculated QNMs from the bound states from the analytical ones, $\log|\omega_n^{(N_\mathrm{t})}-\omega_n|$, as a function of the truncation order $N_{\mathrm{t}}$ of the Taylor expansion, where $\omega_n^{(N_\mathrm{t})}$ is the frequency obtained by applying the mapping to the Taylor expansion given by Eq.~\eqref{TaylorEn} truncated at order $N_{\mathrm{t}}$ and $\omega_n$ is the frequency calculated directly from Eq.~\eqref{QNM_PT}.  
The top and bottom rows correspond to the values from Tabs.~\ref{TabPT1} ($b=V_0=1$) and~\ref{TabPT2} ($b=1/3, V_0=1$), respectively. 
The left and right columns use $b$ and $\alpha=1/b$ as expansion parameters, respectively. 
Overall, appropriate choices of parameters can improve the convergence of the mapping method, whereas higher expansion orders cannot improve it efficiently.
}
\label{figPTconvergence}
\end{figure}

The numerical results illustrate the importance of the expansion parameter in implementing Eq.~\eqref{TaylorEn}.
Fig.~\ref{figPTconvergence} compares the absolute errors $|\omega_n^{(N)}-\omega_n|$ on a logarithmic vertical scale, where $\omega_n^{(N)}$ denotes the frequency obtained by truncating the Taylor expansion at order $N$, while $\omega_n$ is the corresponding analytic result.
Here, $N$ denotes the truncation order and should be distinguished from the overtone index $n$.
The complete numerical results, including the bound-state energies and the frequencies obtained at several truncation orders, are provided in Tabs.~\ref{TabPT1} and~\ref{TabPT2} of Appx.~\ref{app0}.

For $b=V_0=1$, the condition Eq.~\eqref{condPT} is satisfied, and the expansion in $\alpha=1/b$ reproduces the analytic spectrum with increasing accuracy.
For example, the fundamental frequency is $\omega_0=\sqrt{3}/2-i/2\simeq0.866025-0.5i$, whereas the fourth- and tenth-order approximations give $0.868071-0.494574i$ and $0.865901-0.500046i$, respectively.
The improvement also extends to the higher overtones displayed in Appx.~\ref{app0}.
For instance, the tenth-order absolute error for $n=50$ is approximately $3.0\times10^{-4}$.
The logarithmic error curves show an overall decrease with increasing truncation order, although individual successive approximations need not improve monotonically.
In contrast, the expansion in $b$ fails to approach the analytic frequencies; this is because $\pi^{-1}(b)$ lies outside the domain of convergence.

For $b=1/3$ and $V_0=1$, changing the expansion variable to $\alpha$ no longer restores convergence, since the mapped point lies outside the convergence disk for both parameterizations.
In this case, Eq.~\eqref{QNM_PT2} gives the purely imaginary fundamental frequency $\omega_0\simeq-0.381966i$, while the fourth-order expansion in $\alpha$ yields $0.450137-0.962501i$.
Increasing the truncation order does not produce a stable approach to the analytic result, as illustrated by the corresponding logarithmic error curves.
Nonetheless, the exact analytic continuation still yields the QNM spectrum in both parameter regimes.

Besides, as noted in~\cite{agr-qnm-Poschl-Teller-19}, Eq.~\eqref{OmegaPT} implies that there is only a finite number
\bqn
n\le n_\mathrm{max} = \left[\lambda\right],\label{boundN}
\eqn
of bound states for a potential well that vanishes sufficiently rapidly as $x\to \pm\infty$, whereas the QNM spectrum contains infinitely many overtones. 
This might suggest that an infinite set of QNMs cannot be generated from a finite set of bound states.  
However, by analytically continuing the formula for $E_n$ to values $n>n_\mathrm{max}$, which no longer correspond to eigenvalues of a Hermitian problem, one nevertheless recovers the full QNM spectrum.

In the following sections, we further assess the validity of the mapping in the context of spectral instability by elaborating in more detail on the bound states and their analytic continuation to QNMs for a few analytically accessible effective potentials.

\section{QNMs in perturbed delta-function potential barrier}\label{section3}

In this section, we investigate the bound states of a perturbed delta-function potential well and the QNMs of the corresponding potential barrier.

For a delta-function potential barrier,
\begin{eqnarray}
{V}_\mathrm{D1} 
= V_1\delta(C_1 x)  ,
\label{Veff_D1}
\end{eqnarray}
its corresponding potential well obtained by the transformations\footnote{In fact, both choices $\pi(C_1) = \pm i C_1$ serve the purpose.}
\begin{eqnarray}
\pi(V_1) &=& V_1, \nb\\
\pi(C_1) &=& i C_1, \label{piD1}
\end{eqnarray}
has exactly one bound state, with the energy
\bqn
E_0 = -\frac{V_1^2}{4 C_1^2} .
\eqn
Therefore, Eq.~\eqref{MapQNM}, with the minus sign, gives rise to a unique QNM~\cite{agr-qnm-echoes-22}:
\bqn
\omega_0 = -i\frac{V_1}{2 C_1}.\label{omega0D1}
\eqn
Note that the sign of $\pi(C_1)$ in Eq.~\eqref{piD1} is arbitrary since the delta function is even and $C_1$ is squared in the expression for $E_0$. 

To proceed, we add a small perturbation to Eq.~\eqref{Veff_D1}.
It is well known that the following perturbed delta-function potential
\begin{eqnarray}
{V}_\mathrm{D2} 
= V_1\delta(C_1 x) + V_2\delta(C_2 (x-L)) 
\label{Veff_D2}
\end{eqnarray}
with $V_2\ll V_1$ is subject to spectral instability and giving rise to echo modes~\cite{agr-qnm-echoes-22, agr-qnm-instability-065, agr-qnm-echoes-50}.
For the corresponding bound-state energy problem with attractive potential well $\widetilde{V}_\mathrm{D2}=-V_\mathrm{D2}$, we write $E_n=-k_n^2$ with $k_n<0$ and impose exponential decay at both infinities.
The wavefunction is continuous at $x=0$ and $x=L$, while integration of the Schr\"odinger equation gives
\begin{align*}
\widetilde{\Psi}'(0^+)-\widetilde{\Psi}'(0^-)&=-\frac{V_1}{C_1}\widetilde{\Psi}(0),\\
\widetilde{\Psi}'(L^+)-\widetilde{\Psi}'(L^-)&=-\frac{V_2}{C_2}\widetilde{\Psi}(L).
\end{align*}
For $C_1,C_2>0$ and $L>0$, eliminating the amplitudes of the exponential solutions yields the bound-state condition
\bqn
\left(2k_n+\frac{V_1}{C_1}\right)\left(2k_n+\frac{V_2}{C_2}\right)-\frac{V_1V_2}{C_1C_2}\exp\left(-2k_n L\right) = 0,\label{eqBSE}
\eqn
where ${\rm{Re}}(k_n)<0$.

On the other hand, the QNMs are determined by the roots of the following equation~\cite{agr-qnm-echoes-50}:
\bqn
\frac{V_1 V_2}{4C_1C_2\omega_n^2}+ \exp\left(-2i\omega_n L\right)\left(1+i\frac{V_1}{2C_1\omega_n}\right)\left(1+i\frac{V_2}{2C_2\omega_n}\right) =0 .\label{eqQNM}
\eqn

It is reassuring that Eqs.~\eqref{eqBSE} and~\eqref{eqQNM} are directly related by the transformation
\begin{eqnarray}
\pi(L) &=& i L, \nb\\
\pi(V_1) &=& V_1, \nb\\
\pi(V_2) &=& V_2, \nb\\
\pi(C_1) &=& i C_1, \nb\\
\pi(C_2) &=& i C_2, \label{piD2}
\end{eqnarray}
which satisfies Eq.~\eqref{Vcondi}, and $\omega_n = k_n(\pi^{-1}(\{P\}))$.
However, one encounters the same dilemma as in the case of the P\"oschl-Teller effective potential.
On the one hand, since Eq.~\eqref{eqBSE} corresponds to an exponentially decreasing curve intersecting a quadratic one, it admits at most two real roots for $k_n$.
On the other hand, Eq.~\eqref{eqQNM} yields an infinite number of QNMs, often referred to as echo modes, which lie primarily along the real frequency axis and have the following asymptotic form~\cite{agr-qnm-echoes-50}:
\bqn
\omega_n \simeq \left(n+\frac12\right)\frac{\pi}{L} - i\frac{1}{L}\left[\ln\left(\left(n+\frac12\right)\frac{\pi}{L}\right)+\ln 2-\ln\sqrt{\frac{V_1V_2}{C_1C_2}}\right] .\label{exEstiQNM}
\eqn
Therefore, the number of QNMs does not match that of the bound states.
However, if one is allowed to seek roots of Eq.~\eqref{eqBSE} in the complex plane, we recover an infinite number of bound states corresponding to the QNMs in Eq.~\eqref{exEstiQNM}, as demonstrated below in Tab.~\ref{TabDelta2}.
One can further infer the asymptotic behavior of $k_n$ in the complex plane.
In particular, owing to the exponential form, we consider a shift $i n\pi/L$ to the imaginary part of $k_n$, the argument of the exponential remains unchanged.
However, the first term on the left-hand side of Eq.~\eqref{eqBSE} implies that the magnitude of the resulting complex number increases due to the above shift, and the increase is asymptotically quadratic for large overtone number $n$.
This can be readily compensated by slightly decreasing the real part of $k_n$ so that the equality of Eq.~\eqref{eqBSE} holds.
More specifically, $k_n$ asymptotically follows the form
\bqn
k_n \simeq  -\frac{1}{L}\left[\ln\left(\left(n+\frac12\right)\frac{\pi}{L}\right)+\ln 2-\ln\sqrt{\frac{V_1V_2}{C_1C_2}}\right] - i \left(n+\frac12\right)\frac{\pi}{L}
 .\label{exEstiBES}
\eqn
where the real part of $k_n$ decreases asymptotically following the form $-(\ln n)/L$, and the imaginary part increases linearly in $n /L$.
Intriguingly, this result is reminiscent of the consequence of a flip of the QNM frequencies given by Eq.~\eqref{exEstiQNM} along the diagonal direction of the complex frequency plane.
Unfortunately, the price to pay is that these complex bound states are merely an outcome of analytic continuation and do not correspond to any physical states, as the associated wave functions are no longer convergent at the bound.

\begin{figure}[htp]
\centering
\includegraphics[scale=0.49]{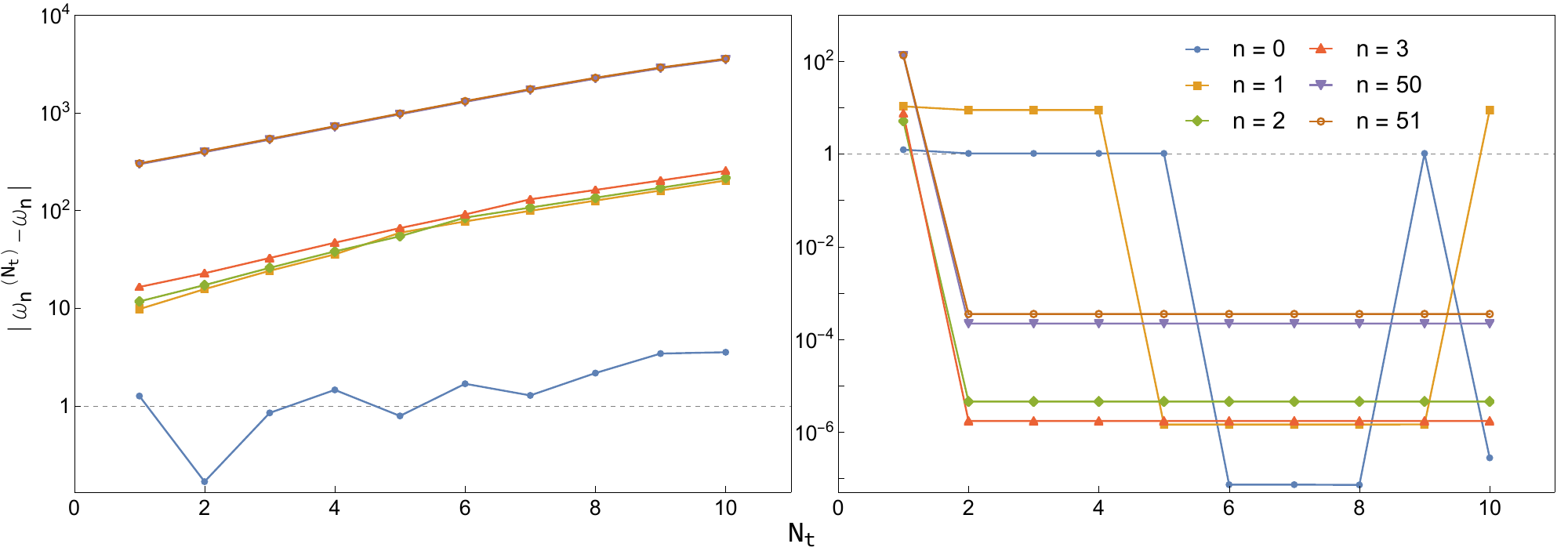}
\caption{The deviations of the numerically calculated QNMs from the bound states from those obtained directly, $\log|\omega_n^{(N_\mathrm{t})}-\omega_n|$, as a function of the truncation order $N_{\mathrm{t}}$ of the Taylor expansion, where $\omega_n^{(N_\mathrm{t})}$ is the frequency obtained by applying the mapping to the Taylor expansion given by Eq.~\eqref{TaylorEn} truncated at order $N_{\mathrm{t}}$ and $\omega_n$ is the frequency calculated directly from Eq.~\eqref{eqQNM}.  
The numerical results corresponds to those presented in Table~\ref{TabDelta2}.
The left and right panels show the results obtained by expanding in the original parameters and their inverses, respectively.
In general, the numerical approach exhibits convergence.
In a few specific cases (e.g., for the purely imaginary modes $n=0,1$ at particular expansion orders $N_{\mathrm{t}}=9,10$), the computed frequencies differ from the corresponding QNMs by an overall sign, which is believed to be a numerical artifact.
}
\label{figDeltaconvergence}
\end{figure}

The perturbed delta-function potential provides a further test of Eq.~\eqref{TaylorEn}, since neither the bound-state energies nor the QNM frequencies are available in closed form.
We consider $V_1=1$, $V_2=0.01$, $C_1=C_2=1$, and $L=1$, and obtain the reference frequencies by directly solving Eq.~\eqref{eqQNM}.
The derivatives required for the Taylor expansion are evaluated by analytically differentiating the implicit relation Eq.~\eqref{eqBSE} and numerically solving the resulting equations, with higher-order derivatives determined iteratively.
The complete results are collected in Tab.~\ref{TabDelta2} of Appx.~\ref{app0}, while Fig.~\ref{figDeltaconvergence} displays the absolute frequency errors as functions of the truncation order on a logarithmic vertical scale.

The comparison again reveals a strong dependence on the choice of expansion parameters.
Expanding in $C_1$, $C_2$, and $L$ does not yield a convergent sequence of approximations to the reference spectrum.
For example, for the mode labeled $n=2$, the absolute error increases from approximately $11.7$ at first order to $215$ at tenth order.
By contrast, the expansion in $1/C_1$, $1/C_2$, and $1/L$ rapidly reproduces the reference frequencies.
Already at fourth order, it yields $\omega_2\simeq3.84709-4.77715i$ and $\omega_{50}\simeq158.601-8.06334i$, in agreement with the direct calculation to the precision displayed in the Appx.~\ref{app0}.
These values remain unchanged at the displayed precision in the fifth- and tenth-order results.
The logarithmic plots make this contrasting behavior explicit: the reciprocal-parameter expansion exhibits a substantial reduction in error, whereas increasing the order of the direct-parameter expansion does not recover the reference spectrum.
When evaluating these errors, the square-root branch in Eq.~\eqref{MapQNM} is chosen consistently with the damped-mode condition $\mathrm{Im}\,\omega_n<0$.

The above results can be understood by examining the radius of convergence of the roots of Eq.~\eqref{eqBSE} with respect to its parameters.
Since Eq.~\eqref{eqBSE} does not contain a singular point, the radius of convergence is determined by the neighborhood in which the implicit function theorem breaks down~\cite{book-methods-mathematical-physics-Hille}.
Specifically, for a given root $k_n$, one seeks in the parameter space the branching points $\{\widetilde{C}_1, \widetilde{C}_2, \widetilde{L}\}$ determined by the following conditions, 
\begin{equation}
\left(2k_n+\frac{V_1}{C_1}\right)\left(2k_n+\frac{V_2}{C_2}\right)-\frac{V_1V_2}{C_1C_2}\exp\left(-2 k_n L\right) = 0 ,\tag{\ref{eqBSE}}
\end{equation}
and the derivative of the above equation with respect to $k_n$
\bqn
\left(2k_n+\frac{V_1}{C_1}\right)+\left(2k_n+\frac{V_2}{C_2}\right)+\frac{LV_1V_2}{C_1C_2}\exp\left(-2 k_n L\right) = 0 .\label{eqBSEprime}
\eqn

Although the above equations do not admit analytic solutions, one can show that the mapping in Eqs.~\eqref{piD2} lies outside the neighborhood considered above.
Subsequently, one faces a similar challenge to justify the validity of the analytic continuation, as in the case of P\"oschl-Teller effective potential.
To see this, we note that Eqs.~\eqref{eqBSE} and~\eqref{eqBSEprime} imply
\bqn
\left(2k_n+\frac{V_1}{C_1}\right)+\left(2k_n+\frac{V_2}{C_2}\right) + L\left(2k_n+\frac{V_1}{C_1}\right)\left(2k_n+\frac{V_2}{C_2}\right)  = 0 .\label{eqBSE2}
\eqn
For simplicity, we consider the asymptotic root governed by Eq.~\eqref{exEstiBES}. 
We focus on the case where the mapped parameters $C_i\to -i C_i$ lie outside the neighborhood of convergence implied by $\{\widetilde{C}_1, \widetilde{C}_2, \widetilde{L}\}$.
Specifically, Eq.~\eqref{eqBSE2} has solution in the limit 
\bqn
\widetilde{C}_1, \widetilde{C}_2, \widetilde{L} \to 0 , \label{limitBP}
\eqn
so that
\bqn
k_n\ll -\frac{V_1}{\widetilde{C}_1}, -\frac{V_2}{\widetilde{C}_2} ,\nb
\eqn
and 
\bqn
\widetilde{L} \simeq -\frac{\widetilde{C}_1 V_2+\widetilde{C}_2 V_1}{V_1 V_2} .\nb
\eqn
Indeed, under the above limit, Eq.~\eqref{eqBSE} approaches
\begin{equation}
\frac{V_1}{C_1}\frac{V_2}{C_2}-\frac{V_1V_2}{C_1C_2} \sim 0 .\nb
\end{equation}
Subsequently, since the radius of convergence is determined by the shortest distance between the expansion point and the nearest branching or singular points, one concludes that
\bqn
\mathcal{R}(C_1)& \le & C_1, \nb\\
\mathcal{R}(C_2)& \le & C_2, \nb\\
\mathcal{R}(L)& \le & L ,
\eqn
and the mapped parameters $\widetilde{C}_1$, $\widetilde{C}_2$, and $\widetilde{L}$ governed by Eqs.~\eqref{piD2} apparently sit outside the above neighborhood.
On the contrary, the branching points in Eq.~\eqref{limitBP} are pushed to infinity for the parameters $1/C_1$, $1/C_2$, and $1/L$, and therefore do not affect the validity of the Taylor expansion.

Lastly, let us make an attempt to analyze the problem from the viewpoint that the QNMs of the effective potential Eq.~\eqref{Veff_D2} are an analytic continuation of the perturbed bound-state energy of the corresponding potential well of Eq.~\eqref{Veff_D1}.
As $V_2\ll V_1$, Eq.~\eqref{eqBSE} gives rise to the following approximate form
\bqn
2k_n\left(2k_n+\frac{V_1}{C_1}\right)-\frac{V_1V_2}{C_1C_2}\exp\left(-2k_n L\right) = 0,\nb
\eqn
\bqn
&&{E_n}'\simeq -\left[\frac{V_1}{4C_1}+\frac14\frac{V_1}{C_1}\sqrt{1+4\exp\left(-2 k_n L\right)\frac{C_1V_2}{V_1C_2}}\right]^2 \simeq-\left[\frac{V_1}{2C_1} + \frac{V_2}{2C_2}\exp\left(-2 k_n L\right)\right]^2  .
\eqn

As an estimation, one plugs in $k_0=-\frac{V_1}{2C_1}$ to the r.h.s. of the above equation and performs the transformation of Eq.~\eqref{piD2}, and finds
\bqn
{\omega_0}' = -\left[\frac{iV_1}{2C_1} + \frac{iV_2}{2C_2}\exp\left(\frac{V_1 L}{C_1}\right)\right] .\label{omega0D2}
\eqn
When compared against Eq.~\eqref{omega0D1}, the fundamental mode governed by Eq.~\eqref{omega0D2} is slightly shifted along the imaginary frequency axis by a small amount $\Delta\omega_0\sim \frac{iV_2}{2C_2}\exp\left(\frac{V_1 L}{C_1}\right)$.
For the specific parameters adopted in Tab.~\ref{TabDelta2}, we have $\Delta\omega_0 \simeq 0.0135914i$ for the fundamental mode.
This readily agrees with the numerical values obtained in the table, where the deviation is ${\Delta\omega_0}' = 0.01373i$.

\section{Bound States of the modified P\"oschl-Teller effective potential}\label{section4}

We consider the modified P\"oschl-Teller potential well with a small discontinuity at $x_c$~\cite{agr-qnm-Poschl-Teller-06, agr-qnm-Poschl-Teller-07, agr-qnm-Poschl-Teller-16, agr-qnm-Poschl-Teller-17},
\begin{eqnarray}
\widetilde{V}_\mathrm{PT} 
= \begin{cases}
   -\frac{V_\mathrm{0-}}{\cosh^2\left(\frac{x}{b}\right)}, &  x \le x_{c}, \\
   -\frac{V_\mathrm{0+}}{\cosh^2\left(\frac{x}{b}\right)}, &  x  > x_{c}.
\end{cases} 
\label{MPT}
\end{eqnarray}
Taking $V_{0-}=V_0$, we assume
\bqn
\frac{|V_{0+}-V_{0-}|}{V_0} \ll 1 .\label{conPer}
\eqn
The perturbed energies are evaluated to first order as $E_n'=E_n+\Delta E_n$, with $\Delta E_n=\langle\Psi_n|H'|\Psi_n\rangle$ and $E_n$ given by Eq.~\eqref{EnPT}.
The perturbation, normalized wavefunctions, and detailed evaluation of the matrix elements are presented in Appx.~\ref{app1}.

It is convenient to describe the position of the discontinuity using
\begin{eqnarray}
y=2\tanh^2\left(\frac{x}{b}\right)-1 ,
\end{eqnarray}
so that $x_c=0$ corresponds to $y_c=-1$, whereas $x_c\to+\infty$ corresponds to $y_c\to+1$.
At the origin, the energy shift receives only the contribution
\begin{eqnarray}
T_{E}^1=T_{O}^1=(V_0-V_\mathrm{0+})\frac{\lambda-n}{2\lambda+1}.
\label{T^1}
\end{eqnarray}
Here, $\Delta E_n=T_E^1$ for even parity and $\Delta E_n=T_O^1$ for odd parity.
For a discontinuity away from the origin, the position-dependent contribution is determined by the integrals in Eqs.~\eqref{rT2Ien0},~\eqref{rT2Ie},~\eqref{rT2Ion1}, and~\eqref{rT2Io}, derived in Appx.~\ref{app1}.
In particular, the correction to the fundamental state is
\begin{eqnarray}
\Delta E_0 
&=&(V_0-V_\mathrm{0+})\left[\frac{\lambda}{2\lambda+1}-2^{\frac{3}{2}}\frac{\lambda}{\sqrt{\pi}} \frac{\Gamma(\lambda+\frac{1}{2})}{\Gamma(\lambda+1)}{(1+y_c)}^{\frac{1}{2}}{_{2}F_1}\left(\frac{1}{2},-\lambda,\frac{3}{2},\frac{1+y_c}{2}\right)\right] .
\label{DeltaE0}
\end{eqnarray}

For the large even indices used in the subsequent analytic continuation, the limiting expressions near the origin and near spatial infinity are given by Eqs.~\eqref{ninfx0} and~\eqref{ninfxinf}, respectively.
Near the origin, writing $y_c=-1+\epsilon$, the approximation that retains the first and last terms of the Jacobi-polynomial expressions requires
\begin{eqnarray}
n^2\epsilon\ll 1 ,
\label{consty_c-1}
\end{eqnarray}
For $n\gg\lambda$, the resulting correction remains perturbative when $(V_0-V_\mathrm{0+})/n\ll1$.
Near spatial infinity, writing $y_c=1-\epsilon$, the analogous approximation leads to the more restrictive condition $(V_0-V_\mathrm{0+})(\epsilon/n^3)(8/\epsilon)^n\ll1$ for the correction to remain small compared with the unperturbed energy.
The endpoint approximations and the derivation of these conditions are collected in Appx.~\ref{app1}.
These results provide the perturbed energies used in the analytic continuation discussed in the following section.

\section{Analytic continuation and the associated QNMs spectrum}\label{section5}

In what follows, we apply the analytic continuation to the perturbed bound-state energy spectrum derived in the last section.
Using Eq.~\eqref{BSP} and again taking the minus sign in Eq.~\eqref{MapQNM}, we formally have
\bqn
{\omega_n}' = -\sqrt{-{E_n}'\left(\pi^{-1}(b), \pi^{-1}(\lambda)\right)} = -\sqrt{-\left(E_n+\Delta E_n\right)} 
=\omega_n + \Delta\omega_n ,
\label{omega'}
\eqn
where $\omega_n$ is given by Eq.~\eqref{QNM_PT} and
\bqn
\Delta\omega_n = \frac12\frac{\Delta E_n}{\sqrt{-E_n}}
\label{delOme},
\eqn
and $\Delta E_n=\Delta E_n\left(\pi^{-1}(b), \pi^{-1}(\lambda)\right)$.

Specifically, the transformations $\pi^{-1}(P_i)$ from bound states to QNMs are given by Eq.~\eqref{piPT}, while the first-order perturbation to the spectrum is entirely governed by $\lambda$ whose transform is given by Eq.~\eqref{piLambdaPT}.
Plugging the latter into Eqs.~\eqref{ninfx0} and~\eqref{ninfxinf}, one obtains the deformation to the QNM spectrum Eq.~\eqref{QNM_PT}, triggered by the metric perturbation implemented by Eq.~\eqref{MPT}.

When the discontinuity point is placed at the origin, using Eqs.~\eqref{OmegaPT} and ~\eqref{T^1}, we obtain
\bqn
\Delta\omega_n \equiv \Delta\omega_n^{(0)}
=ib\frac{V_0-V_\mathrm{0+}}{4\lambda(\pi^{-1}(b))+2}
=ib\frac{V_0-V_{0+}}{2\sqrt{1-4V_{0}b^2}}=\frac{b(V_0-V_{0+})}{2\sqrt{4V_{0}b^2-1}},\label{DeltaOmega_origin}
\eqn
where the last equality makes use of Eq.~\eqref{piLambdaPT}.
This can be readily compared to the correction to QNM first derived in~\cite{agr-qnm-Poschl-Teller-06, agr-qnm-Poschl-Teller-07} by Skakala and Visser. Using the parameters $\alpha_{+}=\frac{1}{2}\sqrt{1-4V_0b^2}$, $\alpha_{-}=\frac{1}{2}\sqrt{1-4V_{0+}b^2}$ and $b_{*}=b$, we compare Eq.~(7.1) of~\cite{agr-qnm-Poschl-Teller-06} and Eq.~\eqref{QNM_PT} 
\bqn
&&\Delta\omega_n =\frac{i}{b}\left[\frac{\arccos\left(1-2\cos\left(\frac{\pi}{2}\sqrt{1-4V_0b^2}\right)\cos\left(\frac{\pi}{2}\sqrt{1-4V_{0+}b^2}\right)\right)}{2\pi}-\lambda(\pi^{-1}(b))\right]\nb\\
&&\simeq\frac{i}{b}\left[\frac{\arccos\left(1-2\cos\left(\frac{\pi}{2}\sqrt{1-4V_0b^2}\right)\cos\left(\frac{\pi}{2}\sqrt{1-4V_{0}b^2}+\frac{\pi(V_0-V_{0+})b^2}{\sqrt{1-4V_{0}b^2}}\right)\right)}{2\pi}-\lambda(\pi^{-1}(b))\right]\nb\\
&&\simeq\frac{i}{b}\left[\frac{\arccos\left(1-2\cos^2\left(\frac{\pi}{2}\sqrt{1-4V_0b^2}\right)\right)}{2\pi}+\frac{(V_0-V_{0+})b^2}{2\sqrt{1-4V_{0}b^2}}-\lambda(\pi^{-1}(b))\right]\nb\\
&&=ib\frac{V_0-V_{0+}}{2\sqrt{1-4V_{0}b^2}} ,
\eqn
where the last equality arises from the fact that when $\alpha_{+}=\alpha_{-}=\frac{1}{2}\sqrt{1-4V_0b^2}$ and $b_{*}=b$, Eq.~(7.1) of~\cite{agr-qnm-Poschl-Teller-06} could reduces to Eq.~\eqref{QNM_PT}.
This is in full agreement with Eq.~\eqref{DeltaOmega_origin}.
Therefore, one concludes that, in this case, the perturbed QNMs coincide analytically with those obtained via the mapping based on analytic continuation from the perturbed bound states evaluated using perturbation theory.

More generally, one considers when the discontinuity is placed near the origin or spatial infinity, for which the convergence of the perturbation theory is well-controlled. 
At the origin, $y_c\sim -1$, we use Eq~\eqref{OmegaPT} and Eq.~\eqref{ninfx0}, taking into account that $\lambda\to \lambda(\pi^{-1}(b))=\mathrm{Re}\lambda+i\mathrm{Im}\lambda$ and $n\gg\mathrm{Re}\lambda$ 
\bqn
&&\Delta\omega_n
\simeq ib(V_0-V_\mathrm{0+})\left[\frac{1}{4(\mathrm{Re}\lambda+i\mathrm{Im}\lambda)+2}+\frac{2^{n-i\mathrm{Im}\lambda}}{n\pi}(1-y_c^2)(1-y_c)^{i\mathrm{Im}\lambda-n}(1+y_c)^{-\frac{1}{2}}\tan{\left(i\pi\mathrm{Im}\lambda\right)}\right]\nb\\
&&~~~~~~~~=\frac{b(V_0-V_\mathrm{0+})\mathrm{Im}\lambda}{4\mathrm{Re}^2\lambda+4\mathrm{Re}\lambda+4\mathrm{Im}^2\lambda+1}\nb\\
&&~~~~~~~~~~-\frac{2^{n}b(V_0-V_\mathrm{0+})}{n\pi}(1-y_c^2)(1-y_c)^{-n}(1+y_c)^{-\frac{1}{2}}\tanh{(\pi\mathrm{Im}\lambda)}\cos{\left(\mathrm{Im}\lambda\ln{\frac{1}{2}(1-y_c)}\right)}\nb\\
&&~~~~~~~~~~+i\left[\frac{b(V_0-V_\mathrm{0+})(2\mathrm{Re}\lambda+1)}{8\mathrm{Re}^2\lambda+8\mathrm{Re}\lambda+8\mathrm{Im}^2\lambda+2}\right.\nb\\
&&~~~~~~~~~~-\left.\frac{2^{n}b(V_0-V_\mathrm{0+})}{n\pi}(1-y_c^2)(1-y_c)^{-n}(1+y_c)^{-\frac{1}{2}}\tanh{(\pi\mathrm{Im}\lambda)}\sin{\left(\mathrm{Im}\lambda\ln{\frac{1}{2}(1-y_c)}\right)}\right].
\label{Del_omey-1}
\eqn
At spacial infinity, $y_c\sim +1$, using the same strategy and further requiring $n\gg\mathrm{Im}\lambda$, Eq.~\eqref{ninfxinf} yields
\bqn
&&\Delta\omega_n
\simeq ib(V_0-V_\mathrm{0+})\left[\frac{1}{4(\mathrm{Re}\lambda+i\mathrm{Im}\lambda)+2}\right.\nb\\
&&~~~~~~~~~~\left.+\frac{8^{n-i\mathrm{Im}\lambda}}{n\left(i\mathrm{Im}\lambda-n\right)\pi}(1-y_c^2)(1-y_c)^{i\mathrm{Im}\lambda-n}(1+y_c)^{-\frac{1}{2}}\tan{\left(i\pi\mathrm{Im}\lambda\right)}\right]\nb\\
&&~~~~~~~~=\frac{b(V_0-V_\mathrm{0+})\mathrm{Im}\lambda}{4\mathrm{Re}^2\lambda+4\mathrm{Re}\lambda+4\mathrm{Im}^2\lambda+1}\nb\\
&&~~~~~~~~~~+\frac{8^{n}b(V_0-V_\mathrm{0+})}{n^2\pi}(1-y_c^2)(1-y_c)^{-n}(1+y_c)^{-\frac{1}{2}}\tanh{(\pi\mathrm{Im}\lambda)}\cos{\left(\mathrm{Im}\lambda\ln{\frac{1}{8}(1-y_c)}\right)}\nb\\
&&~~~~~~~~~~+i\left[\frac{b(V_0-V_\mathrm{0+})(2\mathrm{Re}\lambda+1)}{8\mathrm{Re}^2\lambda+8\mathrm{Re}\lambda+8\mathrm{Im}^2\lambda+2}\right.\nb\\
&&~~~~~~~~~~\left.+\frac{8^{n}b(V_0-V_\mathrm{0+})}{n^2\pi}(1-y_c^2)(1-y_c)^{-n}(1+y_c)^{-\frac{1}{2}}\tanh{(\pi\mathrm{Im}\lambda)}\sin{\left(\mathrm{Im}\lambda\ln{\frac{1}{8}(1-y_c)}\right)}\right],
\label{Del_omey+1}
\eqn
where $\mathrm{Re}\lambda$ and $\mathrm{Im}\lambda$ can be found from Eq.~\eqref{piLambdaPT}, 
\bqn
&\mathrm{Re}\lambda=-\frac{1}{2},~~\mathrm{Im}\lambda=\frac{1}{2}\sqrt{4V_{0}b^2-1} .
\eqn
If Eqs.~\eqref{rT2Ie} and~\eqref{rT2Io} are not approximated by retaining only the first and last terms, the full expressions for the QNM corrections are given in Eqs.~\eqref{Del_omey_full-1} and~\eqref{Del_omey_full+1}, presented in Appx.~\ref{app2}.

To distinguish the two asymptotic behaviors, we denote $y_c=-1+\epsilon$ and $y_c=1-\epsilon$ with $\epsilon\ll1$ and find at the leading order
\begin{eqnarray}
\Delta\omega_n 
&\simeq& \Delta\omega_n^{(0)}-\frac{b\sqrt{\epsilon}(V_0-V_{0+})}{\pi}\left(\frac{2}{n}+\epsilon\right)\tanh{\left(\frac{\pi}{2}\sqrt{4V_0b^2-1}\right)} \nb\\
&\sim& \Delta\omega_n^{(0)}-\frac{\sqrt{2}b(V_0-V_{0+})}{n^2\pi}\tanh{\left(\frac{\pi}{2}\sqrt{4V_0b^2-1}\right)} , \label{QNM_PT_mod}
\end{eqnarray}
for $y_c=-1+\epsilon$, where in the second line one makes use of Eq.~\eqref{consty_c-1}, and
\begin{eqnarray}
\Delta\omega_n 
\simeq \Delta\omega_n^{(0)}+
   \left(\frac{8}{\epsilon}\right)^n\frac{\sqrt{2}b(V_0-V_{0+})}{n^2\pi}\tanh{\left(\frac{\pi}{2}\sqrt{4V_0b^2-1}\right)}\cos{\left(\frac{\ln \epsilon}{2}\sqrt{4V_0b^2-1}\right)} \nb   \\
   ~~~~~~~~~~+i\left(\frac{8}{\epsilon}\right)^n\frac{\sqrt{2}b(V_0-V_{0+})}{n^2\pi}\tanh{\left(\frac{\pi}{2}\sqrt{4V_0b^2-1}\right)}\sin{\left(\frac{\ln \epsilon}{2}\sqrt{4V_0b^2-1}\right)} ,\label{QNM_PT_mod2}
\end{eqnarray}
for $y_c= 1-\epsilon$.

From Eqs.~\eqref{QNM_PT_mod} and~\eqref{QNM_PT_mod2}, one can infer the main feature of the impact on the QNM spectrum due to metric perturbations as follows.
According to Eq.~\eqref{QNM_PT_mod}, as the perturbation moves away from the origin, its initial effect is primarily reflected in the real part of the quasinormal frequency. 
As the perturbation approaches spatial infinity, Eq.~\eqref{QNM_PT_mod2} indicates that its effect is reflected in both the real and the imaginary parts. 
This is consistent with the behavior of the QNM spectrum evolving from parallel to the imaginary axis to parallel to the real axis as the perturbation moves outward from the origin. 

However, based on the result of the first-order correction, the mapping for a metric perturbation placed near spatial infinity yields a result that differs quantitatively from the echo modes, whose asymptotic behavior is known analytically.
Using Eq.~\eqref{QNM_PT_mod2}, the displacement of the high overtones satisfies
\begin{eqnarray}
\frac{\mathrm{Re}(\Delta\omega_{n+2})}{\mathrm{Re}(\Delta\omega_{n})}\propto\frac{1}{\epsilon^2}\gg1,~~~~~~\frac{\mathrm{Im}(\Delta\omega_{n+2})}{\mathrm{Im}(\Delta\omega_{n})} \propto\frac{1}{\epsilon^2}\gg1 .
\end{eqnarray}
As further elaborated in the following section, this behavior differs quantitatively from that of echo modes, which tend to lie closer to the real-frequency axis, with the spacing between adjacent modes remaining finite.

\section{Numerical results and comparison with other approaches}\label{section6}

\begin{figure}[htp]
\centering
\includegraphics[scale=0.56]{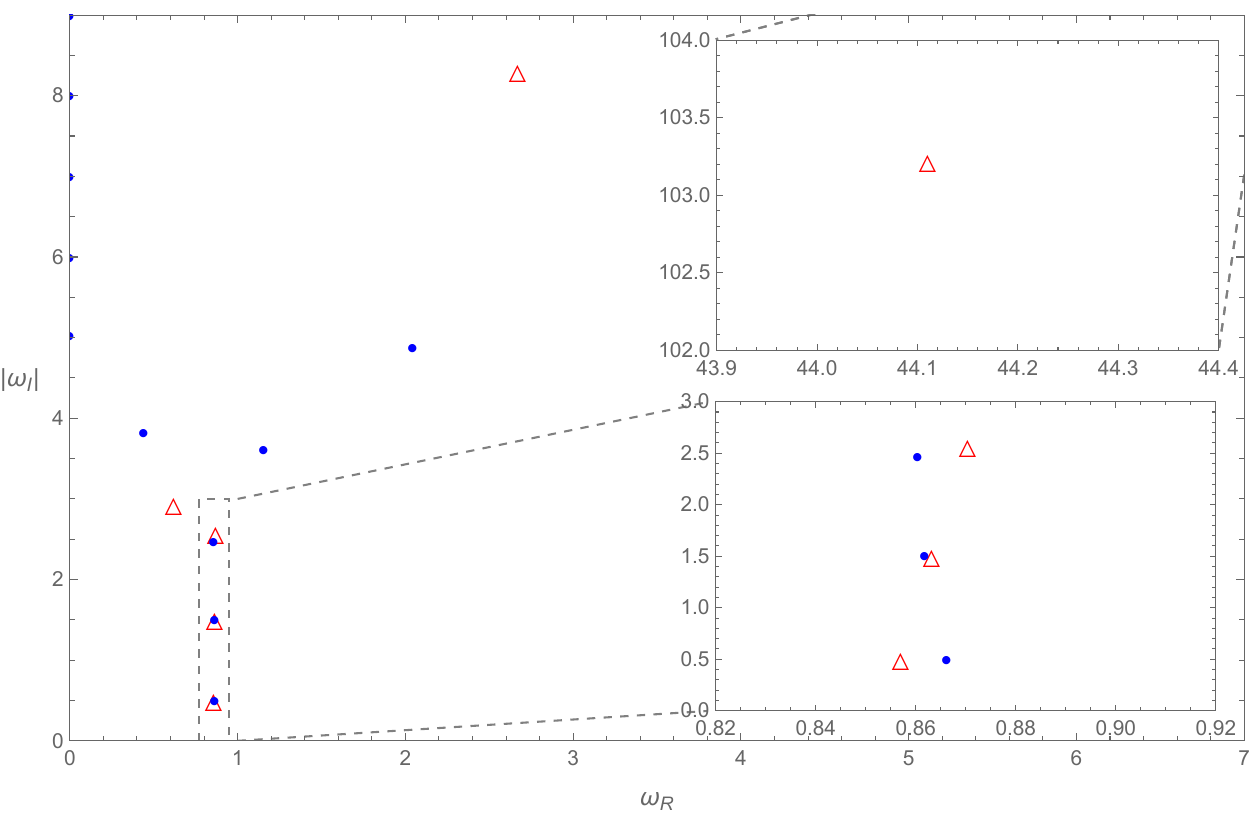}
\caption{QNMs of the modified P\"oschl-Teller potential with $V_0=1$, $V_{0+}=0.99$, $b=1$, and the discontinuity at $y_c=0.62$.
Filled blue dots denote the matrix-method results, checked using $N_\mathrm{grid}=80$, $81$, and $82$.
Empty red triangles denote the frequencies obtained by analytic continuation of the first-order bound-state corrections.
The vertical axis shows $|\mathrm{Im}\omega|$.
The lower inset enlarges the low-lying modes to display the differences between the two approaches.
The upper inset shows the mapped estimate $\omega_6'\simeq44.11-103.22i$, which lies outside the main plotting range.}
\label{fig_per_PT}
\end{figure}

We compare the QNMs obtained by analytic continuation of the first-order bound-state energies with those calculated directly using the matrix method~\cite{agr-qnm-lq-matrix-02, agr-qnm-lq-matrix-03, agr-qnm-lq-matrix-06}.
Fig.~\ref{fig_per_PT} shows the modified P\"oschl-Teller potential with $V_0=1$, $V_{0+}=0.99$, $b=1$, and a discontinuity at $y_c=0.62$.
The mapped frequencies are represented by empty red triangles, while the matrix-method results are shown as filled blue dots.
The latter were checked using three grid sizes, $N_\mathrm{grid}=80$, $81$, and $82$, to assess numerical consistency.
Comparisons with the available semi-analytic results in the appropriate limiting cases provide additional support for using the matrix method as the numerical reference~\cite{agr-qnm-Poschl-Teller-06, agr-qnm-Poschl-Teller-07, agr-qnm-Poschl-Teller-17}.
The figure illustrates three related features: a leftward displacement of the modes, departure from the approximately vertical spectral branch, and increasing disagreement between the two approaches.

First, the unperturbed spectrum given by Eq.~\eqref{QNM_PT} lies on the vertical line $\mathrm{Re}\omega_n=\sqrt{3}/2\simeq0.8660$.
The low-lying perturbed modes remain close to this line, while several are displaced towards smaller real frequencies.
The displacement becomes more pronounced for some of the subsequent overtones.
Because the position-dependent corrections depend on $n$, this behavior is not a uniform translation of the entire spectrum.
For a discontinuity away from the origin, the real-frequency shift is governed by the full corrections in Appx.~\ref{app2}, rather than by the common contribution $\Delta\omega_n^{(0)}$ alone.

Second, the matrix-method spectrum departs from the approximately vertical arrangement as the overtone number increases.
One branch approaches the imaginary axis, whereas another extends towards larger real frequencies.
This bifurcation is consistent with the spectral structure discussed in Ref.~\cite{agr-qnm-Poschl-Teller-17}.
Although the relative change in the potential amplitude is only one percent, the higher modes no longer form a weakly displaced version of the original vertical sequence.
The figure thus illustrates the distinction between a small perturbation of the potential and a small deformation of its QNM spectrum.

Third, the disagreement between the mapped frequencies and the matrix-method results becomes pronounced in the region where the spectrum departs from the original branch.
The lower inset shows that differences are already present among the low-lying modes, even though both calculations place them near the unperturbed sequence.
For higher overtones, the mapped corrections grow substantially and no longer reproduce the numerical frequencies.
The upper inset illustrates this behavior through the mapped estimate $\omega_6'\simeq44.11-103.22i$, compared with the unperturbed frequency $\omega_6\simeq0.8660-6.5i$.
Such a correction lies outside the small-correction regime assumed in the linearization leading to Eq.~\eqref{delOme}.
The mapped results therefore indicate strong spectral sensitivity without providing a quantitatively reliable description of the resulting branches.

The association between the bifurcation and the growing disagreement can be understood in terms of the perturbative construction.
The mapped spectrum is obtained from $E_n'=E_n+\Delta E_n$ by retaining only the first-order frequency correction in Eq.~\eqref{delOme}.
Smallness of $|V_{0+}-V_0|/V_0$ and convergence of the bound-state matrix elements do not ensure that this approximation remains accurate after analytic continuation for all overtone numbers.
The continued coefficients can be strongly amplified as $n$ increases, as illustrated by the factor $(8/\epsilon)^n$ in the near-infinity estimate Eq.~\eqref{QNM_PT_mod2}.
Although this limiting expression does not determine the bifurcation point in Fig.~\ref{fig_per_PT}, it demonstrates why a small potential perturbation need not produce a small mapped frequency correction.
The departure from the original spectral branch and the loss of quantitative agreement can therefore occur in the same overtone range.

This interpretation concerns the accuracy of the first-order approximation.
The calculation does not provide higher-order error estimates or establish an exact coincidence between the bifurcation and a convergence boundary.
In the Green's-function description, the QNMs are poles, and the bifurcation represents a reorganization of their pattern.
A nearby singularity or branch point in the dependence of the frequencies on complexified potential parameters could restrict a perturbative expansion, but such a point has not been identified here.
The figure alone therefore does not establish a new singularity of the Green's function at the bifurcation.

The perturbed delta-function example in Sec.~\ref{section3} provides a complementary perspective.
In that case, the full secular equations, Eqs.~\eqref{eqBSE} and~\eqref{eqQNM}, are related by analytic continuation, and the selected QNMs are recovered using a convergent expansion in suitable parameters.
This comparison emphasizes the distinction between continuing the full spectral problem and continuing a first-order approximation to its solutions.
For the modified P\"oschl-Teller potential, the numerical comparison establishes the limited accuracy of the present truncation as the spectral deformation grows.
Determining whether higher-order corrections restore agreement, or whether the continuation encounters a convergence-limiting singularity, requires an analysis beyond the first-order calculation presented here.

\section{Further discussions and concluding remarks}\label{section6}

To summarize, we investigated the analytic continuation scheme that maps the bound-state spectrum of a potential well to the QNM spectrum of the corresponding potential barrier in the context of spectral instability.
As an analytically accessible example, we studied the perturbed delta-function potential barrier.
Although the equations governing the bound states and QNMs are formally related through analytic continuation, the number of bound states does not match the number of QNMs.
This discrepancy can be formally resolved by allowing complex bound-state energies, but the resulting states no longer correspond to physical bound states because the associated wavefunctions diverge at the boundary.

We then analyzed the modified P\"oschl-Teller effective potential containing a small discontinuity.
The bound-state spectrum and the corresponding perturbative corrections were derived analytically using first-order Rayleigh–Schr\"odinger perturbation theory.
When the discontinuity is placed near the maximum of the effective potential, the perturbative bound-state corrections can be analytically continued to yield quasinormal frequencies that agree with known numerical and semi-analytic results.
However, when the discontinuity is moved toward spatial infinity, although the perturbation theory for the bound states remains well controlled, the analytic continuation produces a spectrum that deviates significantly from the actual QNMs, particularly for the asymptotically high overtones.
This behavior suggests that, in the presence of spectral instability, where small deformations of the effective potential lead to large changes in the quasinormal spectrum, the mapping based on analytic continuation encounters significant challenges.
Nonetheless, this does not necessarily invalidate the approach, as higher-order corrections or a more suitable perturbative framework may alleviate the problem and lead to improved agreement.

The present analysis highlights several conceptual issues in the mapping between bound states and QNMs.
First, the number of bound states and QNMs generally does not coincide, and the mapping requires analytic continuation to complex energies at the expense of losing physically acceptable bound-state wavefunctions.
Second, the mathematical justification of the analytic continuation itself remains subtle, since the mapped parameters typically lie outside the naive radius of convergence.
Third, although perturbation theory for the bound-state problem may remain valid, this does not guarantee that the analytically continued spectrum faithfully reproduces the QNMs, at least at the lowest order.
Fourth, numerically, for the scheme proposed by V\"olkel to work, the metric parameters must be carefully chosen so that their transformations are carried out in a region where the underlying Taylor expansion is convergent.
Our findings indicate that in regimes where spectral instability becomes prominent, the analytic continuation between bound states and QNMs is a subtle issue that warrants a more systematic analysis and motivates further studies of its limitations, possible extensions, and physical implications.

\section*{Acknowledgements}

We gratefully acknowledge the financial support from Brazilian agencies 
Funda\c{c}\~ao de Amparo \`a Pesquisa do Estado de S\~ao Paulo (FAPESP), 
Funda\c{c}\~ao de Amparo \`a Pesquisa do Estado do Rio de Janeiro (FAPERJ), 
Conselho Nacional de Desenvolvimento Cient\'{\i}fico e Tecnol\'ogico (CNPq), 
and Coordena\c{c}\~ao de Aperfei\c{c}oamento de Pessoal de N\'ivel Superior (CAPES).
This work is supported by the National Natural Science Foundation of China (NSFC).
A part of this work was developed under the project Institutos Nacionais de Ci\^{e}ncias e Tecnologia - Física Nuclear e Aplica\c{c}\~{o}es (INCT/FNA) Proc. No. 408419/2024-5.
This research is also supported by the Center for Scientific Computing (NCC/GridUNESP) of São Paulo State University (UNESP).

\appendix

\section{The bound-state energies and the QNMs obtained for a few effective potentials}\label{app0}

In this Appendix, we show the bound-state energies and the corresponding QNMs obtained for the P\"oschl-Teller and the perturbed delta-function effective potentials.

\begin{table}[htbp]
\centering
\small
\caption{The bound-state energies (complex numbers) and the corresponding QNMs for the P\"oschl-Teller potential barrier obtained using the numerical scheme given by Eq.~\eqref{TaylorEn}, compared against the analytic results. 
The calculations are carried out using the parameters $b=V_0=1$.}
\label{TabPT1}
\setlength{\tabcolsep}{5pt}
\renewcommand{\arraystretch}{1.2}
\begin{tabular}{llccccccc}
\hline\hline
overtone index & & $n=0$ & $n=1$ & $n=2$ & $n=3$ &$\cdots$ & $n=50$ & $n=51$ \\
\hline
bound-state energies & Re & $-0.381966$ & $-0.145898$ & $-1.90983$ & $-5.67376$ &$\cdots$& $-2438.58$ & $-2538.34$ \\
from Eq.~\eqref{EnPT} & Im ($i$) & $0$ & $0$ & $0$ & $0$ &$\cdots$& $0$ & $0$ \\
\hline\hline
QNMs & Re & $0.866025$ & $0.866025$ & $0.866025$ & $0.866025$ &$\cdots$& $0.866025$ & $0.866025$ \\
from Eq.~\eqref{QNM_PT} & Im ($i$) & $-0.5$ & $ - 1.5$ & $-2.5$ & $-3.5$ &$\cdots$& $-50.5$ & $-51.5$ \\
\hline\hline
\multicolumn{9}{c}{QNMs obtained by mapping using Eq.~\eqref{TaylorEn} in terms of the parameter $b$}\\
\hline
& Re & $0.438371$ & $1.14164 $ & $3.04455 $ & $4.88235  $ &$\cdots$& $90.33 $ & $92.1474$ \\
1st order & Im ($i$) & $0.389671$ & $- 0.42705$ & $ - 1.03329$ & $- 1.59846$ &$\cdots$& $  - 27.4853$ & $ - 28.0356$ \\
\hline
 & Re & $0.616181$ & $1.2917$ & $2.85155 $ & $4.39146$ &$\cdots$& $76.319  $ & $77.849 $ \\
4th order & Im ($i$) & $ 0.592041 $ & $ - 5.34612$ & $ - 10.5823$ & $ - 15.8139$&$\cdots$ & $- 261.575$ & $ - 266.804$ \\
\hline
 & Re & $0.621448  $ & $0.745315 $ & $1.28571 $ & $1.84532$ &$\cdots$& $28.595 $ & $29.1646$ \\
5th order & Im ($i$) & $0.738152$ & $7.21329$ & $14.2284$ & $21.2489$&$\cdots$ & $351.331$ & $358.354$ \\
\hline
& Re & $1.07223  $ & $20.1808  $ & $40.4081 $ & $60.6348  $ &$\cdots$& $1011.28$ & $1031.5$ \\
10th order & Im ($i$) & $ 0.716298$ & $ - 9.3998$ & $ - 18.9111$ & $ - 28.4098$ &$\cdots$& $-474.571$ & $ - 484.063$\\ 
\hline\hline
\multicolumn{9}{c}{QNMs obtained by mapping using Eq.~\eqref{TaylorEn} in terms of the parameter $\alpha=1/b$}\\
\hline
& Re & $0.872872$ & $0.474756$ & $1.28168 $ & $2.0697 $ &$\cdots$& $39.0165$ & $39.8026 $ \\
1st order & Im ($i$) & $-0.195699$ & $1.02692$ & $2.45452 $ & $3.77073$ &$\cdots$& $ 63.6333$ & $ 64.9054$ \\
\hline
 & Re & $0.868071  $ & $0.864373$ & $0.862198$ & $0.86115 $ &$\cdots$& $0.858804$ & $0.858801 $ \\
4th order & Im ($i$) & $- 0.494574$ & $- 1.49007$ & $ - 2.48971$ & $ - 3.48984$&$\cdots$ & $ - 50.4909$ & $ - 51.4909$ \\
\hline
 & Re & $0.864155 $ & $0.85756 $ & $0.855295 $ & $0.85441$ &$\cdots$& $0.852984$ & $0.852983 $ \\
5th order & Im ($i$) & $-0.493503$ & $- 1.49189$ & $ - 2.49308$ & $ - 3.49392$&$\cdots$ & $ - 50.4966$ & $ - 51.4966$ \\
\hline
& Re & $0.865901 $ & $0.865879 $ & $0.865905$ & $0.865923 $ &$\cdots$& $0.865976 $ & $0.865977 $ \\
10th order & Im ($i$) & $ - 0.500046$ & $ - 1.50018$ & $ - 2.50022$ & $ - 3.50024$ &$\cdots$& $ - 50.5003$ & $ - 51.5003$ \\
\hline\hline
\end{tabular}
\end{table}

\begin{table}[htbp]
\centering
\small
\caption{The same as Tab.~\ref{TabPT1}, but using the parameters $b=1/3$ and $V_0=1$.}
\label{TabPT2}
\setlength{\tabcolsep}{5pt}
\renewcommand{\arraystretch}{1.2}
\begin{tabular}{llccccccc}
\hline\hline
overtone index & & $n=0$ & $n=1$ & $n=2$ & $n=3$ &$\cdots$ & $n=50$ & $n=51$ \\
\hline
bound-state energies & Re & $-0.0916731$ & $-7.27502$ & $-32.4584$ & $-75.6417$ &$\cdots$& $-22409.3$ & $-23316.4$ \\
from Eq.~\eqref{EnPT} & Im ($i$) & $0$ & $0$ & $0$ & $0$ &$\cdots$& $0$ & $0$ \\
\hline\hline
QNMs & Re & $0$ & $0$ & $0$ & $0$ &$\cdots$& $0$ & $0$ \\
from Eq.~\eqref{QNM_PT2} & Im ($i$) & $-0.38197$ & $-3.38197$ & $-6.38197 $ & $ -9.38197$ &$\cdots$& $-150.382 $ & $ -153.382 $ \\
\hline\hline
\multicolumn{9}{c}{QNMs obtained by mapping using Eq.~\eqref{TaylorEn} in terms of the parameter $b$}\\
\hline
& Re & $0.810385$ & $9.60825 $ & $19.4057  $ & $29.1904  $ &$\cdots$& $488.787$ & $498.566$ \\
1st order & Im ($i$) & $0.847115$ & $ - 8.21591 $ & $ - 16.5192$ & $ - 24.8094 $ &$\cdots$& $ - 414.149 $ & $ - 422.432$ \\
\hline
 & Re & $7.163 $ & $532.629$ & $1065.1 $ & $1597.58$ &$\cdots$& $26624.6$ & $27157.1 $ \\
4th order & Im ($i$) & $3.18898$ & $144.905$ & $289.704$ & $ 434.512$&$\cdots$ & $ 7240.75 $ & $7385.57$ \\
\hline
 & Re & $7.64289$ & $1627.7$ & $3255.46 $ & $4883.21$ &$\cdots$& $81387.4$ & $83015.2$ \\
5th order & Im ($i$) & $ 8.0882 $ & $ - 811.828$ & $ - 1623.76$ & $- 2435.68 $&$\cdots$ & $ - 40595.8$ & $ - 41407.7 $ \\
\hline
& Re & $397.602$ & $344924.$ & $290.71 $ & $1.0\times10^6 $ &$\cdots$& $1.7\times10^7$ & $1.76\times10^7$ \\
10th order & Im ($i$) & $ - 494.752 $ & $ - 500338.$ & $ - 1.0\times10^6 $ & $- 1.5\times10^6 $ &$\cdots$& $-2.5\times10^7$ & $ - 2.55\times10^7$\\ 
\hline\hline
\multicolumn{9}{c}{QNMs obtained by mapping using Eq.~\eqref{TaylorEn} in terms of the parameter $\alpha=1/b$}\\
\hline
& Re & $0.515841$ & $2.24263$ & $4.60051$ & $6.95866$ &$\cdots$& $117.805$ & $120.164$ \\
1st order & Im ($i$) & $ - 0.147869 $ & $ 3.9111$ & $7.74232$ & $11.5634$ &$\cdots$& $190.928$ & $ 194.744 $ \\
\hline
 & Re & $0.450137$ & $0.356424$ & $0.3276$ & $0.315906$ &$\cdots$& $0.29061$ & $0.290575$ \\
4th order & Im ($i$) & $ - 0.962501 $ & $ - 3.6467 $ & $ - 6.61258 $ & $ - 9.60033 $&$\cdots$ & $ - 150.576$ & $ - 153.576 $ \\
\hline
 & Re & $0.178071$ & $0.154836$ & $0.14544$ & $0.141397$ &$\cdots$& $0.132191$ & $0.132178$ \\
5th order & Im ($i$) & $ - 1.1079$ & $ - 3.82246 $ & $ - 6.78232$ & $ - 9.76679 $&$\cdots$ & $ - 150.734 $ & $ - 153.734$ \\
\hline
& Re & $0.45277 $ & $0.310468 $ & $0.280491 $ & $0.268942$ &$\cdots$& $0.244882 $ & $0.24485$ \\
10th order & Im ($i$) & $ 0.805832$ & $ 3.52555$ & $ 6.50388$ & $9.49647$ &$\cdots$& $ 150.482$ &  $153.482 $ \\
\hline\hline
\end{tabular}
\end{table}

\begin{table}[htbp]
\centering
\small
\caption{The bound-state energy and the corresponding QNMs for the perturbed delta-function potential barrier obtained using the numerical scheme given by Eq.~\eqref{TaylorEn}, compared against those obtained by solving Eq.~\eqref{eqQNM}. 
The calculations are carried out for the parameters $V_1=1, V_2=0.01, C_1=C_2=1$, and $L=1$.}
\label{TabDelta2}
\setlength{\tabcolsep}{5pt}
\renewcommand{\arraystretch}{1.2}
\begin{tabular}{llccccccc}
\hline\hline
overtone index & & $n=0$ & $n=1$ & $n=2$ & $n=3$ &$\cdots$ & $n=50$ & $n=51$ \\
\hline
bound-state energies & Re  &$ -0.263919$ & $ -19.5513$ & $-8.02101$ & $26.001$ &  $\cdots$& $25089.3$ & $26095.7$ \\
from Eq.~\eqref{eqBSE} & Im ($i$)  &$ 0$ & $ 0$& $ - 36.7562$ & $ - 75.014 $ &$\cdots$& $ - 2557.71 $ & $- 2614.72 $ \\
\hline\hline
QNMs (echo modes) & Re &$ 0$ & $ 0$& $3.84709 $ & $7.25925$ &$\cdots$& $158.601$ & $161.744$ \\
from Eq.~\eqref{eqQNM} & Im ($i$) &$ -0.51373$ & $ -4.42169$& $ - 4.77715$ & $ - 5.16679$ &$\cdots$& $ - 8.06334$ & $ - 8.08292 $ \\
\hline\hline
\multicolumn{9}{c}{QNMs obtained by mapping using Eq.~\eqref{TaylorEn} in terms of ${C_1}$ ,${C_2}$, and ${L}$}\\
\hline
& Re &$ 0.410039$ & $ 9.70938 $& $12.1971$ & $14.7142$ &$\cdots$& $103.235$ & $105.0$ \\
1st order & Im ($i$) &$ 0.679009$ & $  - 3.29034$& $ 3.43192$ & $ 9.49191$ &$\cdots$& $283.087 $ & $288.787 $ \\
\hline
 & Re &$ 0.0976221$ & $ 9.05789$& $28.3562 $ & $46.8023$ &$\cdots$& $843.31$ & $859.772$ \\
4th order & Im ($i$) &$ 0.940073$ & $  - 38.7134 $& $- 33.8602$ & $ - 29.9869 $ &$\cdots$ & $ 187.22 $ & $191.926$ \\
\hline
 & Re &$ 0.783878$ & $ 7.11639$& $18.3055$ & $42.3878$ &$\cdots$& $1105.06$ & $1127.12$ \\
5th order & Im ($i$) &$  - 0.601616$ & $  54.2257$& $ - 57.1506 $ & $ - 60.8845 $ &$\cdots$ & $ - 166.69$ & $- 168.618$ \\
\hline
& Re &$ 2.34769$ & $ 182.842$& $218.31$ & $254.899$ &$\cdots$& $1748.61$ & $1778.84$ \\
10th order & Im ($i$) &$  - 3.15994$ & $ - 90.1887 $& $ - 19.1961$ & $ 48.4135$ &$\cdots$& $3089.48$ & $3152.86$\\ 
\hline\hline
\multicolumn{9}{c}{QNMs obtained by mapping using Eq.~\eqref{TaylorEn} in terms of ${1}/{C_1}$ ,${1}/{C_2}$, and ${1}/{L}$}\\
\hline
& Re &$ 0.40387$ & $ 3.47612 $& $1.13802 $ & $5.17203$ &$\cdots$& $195.405$ & $199.387$ \\
1st order & Im ($i$) &$ 0.653475$ & $ 5.62447$& $ - 9.10102$ & $ - 12.2791 $ &$\cdots$& $ - 134.941 $ & $  - 137.437 $ \\
\hline
 & Re &$ 0$ & $ 0$& $3.84709$ & $7.25925 $ &$\cdots$& $158.601$ & $161.744$ \\
4th order & Im ($i$) &$ 0.51373 $ & $ 4.42169$& $ - 4.77715$ & $ - 5.16679$&$\cdots$ & $ - 8.06334$ & $ - 8.08292$ \\
\hline
 & Re &$ 0$ & $ 0$&  $3.84709$ & $7.25925$ &$\cdots$& $158.601$ & $161.744$ \\
5th order & Im ($i$) &$ -0.51373 $ & $ -4.42169$& $ - 4.77715$ & $ - 5.16679$ &$\cdots$ & $ - 8.06334$ & $- 8.08292$ \\
\hline
& Re &$ 0$ & $ 0$& $3.84709$ & $7.25925 $&$\cdots$& $158.601$ & $161.744$ \\
10th order & Im ($i$) &$ -0.51373 $ & $ 4.42169$& $ - 4.77715$ & $ - 5.16679$ &$\cdots$& $ - 8.06334$ & $- 8.08292$ \\
\hline\hline
\end{tabular}
\end{table}

\section{Derivation of the bound-state energy corrections}\label{app1}

In this appendix, we present the detailed derivation of the bound-state energy corrections summarized in Sec.~\ref{section4}.
It includes the unperturbed wavefunctions and their normalization, the evaluation of the perturbation matrix elements, and the limiting expressions for a discontinuity near the origin or near spatial infinity.

We first summarize the setup of the perturbation problem, present the wavefunctions and their normalization.
With $V_{0-}=V_0$, the perturbation of the P\"oschl-Teller potential well is
\begin{eqnarray}
H' 
= \begin{cases}
   0, &  x \le x_{c}, \\
   \frac{V_0-V_\mathrm{0+}}{\cosh^2\left(\frac{x}{b}\right)}, &  x  > x_{c}.
\end{cases} 
\label{HP}
\end{eqnarray}
The first-order Rayleigh-Schr\"odinger corrections to the energies and wavefunctions are
\begin{eqnarray}
E'_{n}&=& E_n+\Delta E_n  = E_n+\langle\Psi_n|H'|\Psi_n\rangle,\nb\\
|\Psi'_{n}\rangle&=&|\Psi_n\rangle+\sum_{i\neq n}\frac{\langle\Psi_i|H'|\Psi_n\rangle}{E_n-E_i}|\Psi_i\rangle,
\label{BSP}
\end{eqnarray}
To evaluate the matrix elements, we use the wavefunctions associated with the unperturbed energies in Eq.~\eqref{EnPT} and write
\begin{eqnarray}
\Psi(x)=\frac{N\psi(x)}{{\cosh^{\lambda}\left(\frac{x}{b}\right)}},~~~
\label{WF}
\end{eqnarray}
where $N$ is the normalization constant.
Substitution into Eq.~\eqref{bound_frequency_domain} gives two independent hypergeometric solutions,
\begin{eqnarray}
\psi_1(x)={_{2}F_1}\left(A,B,\frac{1}{2},-\sinh^2\left(\frac{x}{b}\right)\right),
&~&\psi_2(x)=\sinh\left(\frac{x}{b}\right){_{2}F_1}\left(A+\frac{1}{2},B+\frac{1}{2},\frac{3}{2},-\sinh^2\left(\frac{x}{b}\right)\right),\label{SOH}\\
A=\frac{1}{2}(b\Omega-\lambda),&~&B=-\frac{1}{2}(b\Omega+\lambda) ,\nb
\end{eqnarray}
The bound-state solutions are obtained by truncating the hypergeometric series.
For the two parity sectors, this occurs when $A=-n$ or $A+1/2=-n$~\cite{agr-qnm-Poschl-Teller-17}, with $n$ a non-negative integer,
\begin{eqnarray}
\frac{1}{2}(b\Omega-\lambda)=-n,~~~\frac{1}{2}(b\Omega-\lambda)+\frac{1}{2}=-n .
\label{DS}
\end{eqnarray}
Combining the two sectors gives the spectrum in Eq.~\eqref{EnPT}.
Expressed in terms of the energy-level index, the even- and odd-parity solutions become
\begin{eqnarray}
\psi_{E}(x)&=&{_{2}F_1}\left(-\frac{n}{2},\frac{n}{2}-\lambda,\frac{1}{2},-\sinh^2\left(\frac{x}{b}\right)\right),\nb\\
\psi_{O}(x)&=&\sinh\left(\frac{x}{b}\right){_{2}F_1}\left(-\frac{1}{2}(n-1),\frac{1}{2}(n-1)-\lambda+1,\frac{3}{2},-\sinh^2\left(\frac{x}{b}\right)\right) .\label{SOHn}
\end{eqnarray}
where $n=0,2,\cdots$ for even parity and $n=1,3,\cdots$ for odd parity.
At large $|x|$, these functions behave as
\begin{eqnarray}
\psi_{E/O}(x)\sim e^{n\left|\frac{x}{b}\right|}.
\end{eqnarray}
and hence
\begin{eqnarray}
\Psi(x)\sim e^{-(\lambda-n)\left|\frac{x}{b}\right|} ,
\end{eqnarray}
The bound-state boundary condition therefore imposes Eq.~\eqref{boundN}.
As in the perturbed delta-function example discussed in Sec.~\ref{section3}, continuing the expressions beyond the physical bound-state spectrum requires abandoning the bound-state boundary condition.

The truncated hypergeometric functions can be expressed in terms of Jacobi polynomials as
\begin{eqnarray}
{_{2}F_1}\left(-\frac{n}{2},\frac{n}{2}-\lambda,\frac{1}{2},-\sinh^2\left(\frac{x}{b}\right)\right)=
\frac{\Gamma\left(\frac{n}{2}+1\right)\Gamma\left(\frac{1}{2}\right)}{\Gamma\left(\frac{n}{2}+\frac{1}{2}\right)}\cosh^{n}\left(\frac{x}{b}\right)P_{n/2}^{\lambda-n,-1/2}\left(y\right),\nb\\
{_{2}F_1}\left(-\frac{1}{2}(n-1),\frac{1}{2}(n-1)-\lambda+1,\frac{3}{2},-\sinh^2\left(\frac{x}{b}\right)\right)=
\frac{\Gamma\left(\frac{n}{2}+\frac{1}{2}\right)\Gamma\left(\frac{3}{2}\right)}{\Gamma\left(\frac{n}{2}+1\right)}\cosh^{n-1}\left(\frac{x}{b}\right)P_{(n-1)/2}^{\lambda-n,1/2}\left(y\right),\nb\\ \label{FTPn}
\end{eqnarray}
where $y$ is the coordinate introduced in Sec.~\ref{section4}.
Equation~\eqref{FTPn} follows from the identities
\begin{eqnarray}
{_{2}F_1}(A,B,C,z)=(1-z)^{-A}{_{2}F_1}\left(A,C-B,C,\frac{z}{z-1}\right),\nb\\
P_{n}^{\alpha,\beta}(z)=\frac{(\alpha+1)_{n}}{n!}{_{2}F_1}\left(-n,\alpha+\beta+n+1,\alpha+1,\frac{1-z}{2}\right).
\end{eqnarray}
Using the integral relation
\begin{eqnarray}
\int_{-1}^{1} dz (1-z)^{\alpha-1}(1+z)^{\beta}\left(P_{n}^{\alpha,\beta}(z)\right)^2=\frac{2^{\alpha+\beta}\Gamma(\alpha+n+1)\Gamma(\beta+n+1)}{n!\alpha\Gamma(\alpha+\beta+n+1)},
\label{IR1}
\end{eqnarray}
we obtain the normalization constants for the two parity sectors,
\begin{eqnarray}
N_{E}&=&\left(\frac{1}{b}\right)^{\frac{1}{2}}\left[\frac{2(\lambda-n)}{(n+1)(\lambda-\frac{n}{2})}\frac{\Gamma(\lambda-\frac{n}{2}+\frac{1}{2})\Gamma(\frac{n}{2}+\frac{3}{2})}{\Gamma(\frac{1}{2})\Gamma(\lambda-\frac{n}{2})\Gamma(\frac{1}{2})\Gamma(\frac{n}{2}+1)}\right]^{\frac{1}{2}},\nb\\
N_{O}&=&\left(\frac{1}{b}\right)^{\frac{1}{2}}\left[\frac{2(\lambda-n)}{\lambda-\frac{1}{2}(n+1)}\frac{\Gamma(\lambda-\frac{n}{2}+1)\Gamma(\frac{n}{2}+1)}{\Gamma(\frac{3}{2})\Gamma(\lambda-\frac{n}{2}-\frac{1}{2})\Gamma(\frac{1}{2})\Gamma(\frac{n}{2}+\frac{1}{2})}\right]^{\frac{1}{2}}.
\label{NCn}
\end{eqnarray}

Now, we proceed to derive the elements of the energy matrix.
The first-order energy correction in Eq.~\eqref{BSP} can be separated into a contribution from the positive half-line and a position-dependent term,
\begin{eqnarray}
\Delta E_n = \langle\Psi_n|H'|\Psi_n\rangle&=&\int_{x_{c}}^{+\infty}dx\frac{V_0-V_\mathrm{0+}}{\cosh^2\left(\frac{x}{b}\right)}\Psi_n^\dagger \Psi_n\nb\\
&=&\int_{0}^{+\infty}dx\frac{V_0-V_\mathrm{0+}}{\cosh^2\left(\frac{x}{b}\right)}\Psi_n^\dagger \Psi_n-\int_{0}^{x_c}dx\frac{V_0-V_\mathrm{0+}}{\cosh^2\left(\frac{x}{b}\right)}\Psi_n^\dagger \Psi_n .
\label{<H>}
\end{eqnarray}
For even parity, the Jacobi-polynomial representation gives
\begin{eqnarray}
\Delta E_{n/2} 
=(V_0-V_\mathrm{0+})\left[\frac{\int_{-1}^{1}dy(1-y)^{\lambda-n}(1+y)^{-1/2}\left(P_{n/2}^{\lambda-n,-1/2}(y)\right)^2}{\int_{-1}^{1}dy(1-y)^{\lambda-n-1}(1+y)^{-1/2}\left(P_{n/2}^{\lambda-n,-1/2}(y)\right)^2}\right.\nb\\
\left.-\frac{\int_{-1}^{y_c}dy(1-y)^{\lambda-n}(1+y)^{-1/2}\left(P_{n/2}^{\lambda-n,-1/2}(y)\right)^2}{\int_{-1}^{1}dy(1-y)^{\lambda-n-1}(1+y)^{-1/2}\left(P_{n/2}^{\lambda-n,-1/2}(y)\right)^2}\right] .
\label{<H>n/2}
\end{eqnarray}
while for odd parity it gives
\begin{eqnarray}
\Delta E_{(n-1)/2} 
=(V_0-V_\mathrm{0+})\left[\frac{\int_{-1}^{1}dy(1-y)^{\lambda-n}(1+y)^{1/2}\left(P_{(n-1)/2}^{\lambda-n,1/2}(y)\right)^2}{\int_{-1}^{1}dy(1-y)^{\lambda-n-1}(1+y)^{1/2}\left(P_{(n-1)/2}^{\lambda-n,1/2}(y)\right)^2}\right.\nb\\
\left.-\frac{\int_{-1}^{y_c}dy(1-y)^{\lambda-n}(1+y)^{1/2}\left(P_{(n-1)/2}^{\lambda-n,1/2}(y)\right)^2}{\int_{-1}^{1}dy(1-y)^{\lambda-n-1}(1+y)^{1/2}\left(P_{(n-1)/2}^{\lambda-n,1/2}(y)\right)^2}\right].
\label{<H>n-1/2}
\end{eqnarray}
When the discontinuity is at the origin, $y_c=-1$, only the first term contributes in each expression.
The identity
\begin{eqnarray}
\int_{-1}^{1} dz (1-z)^{\alpha}(1+z)^{\beta}\left(P_{n}^{\alpha,\beta}(z)\right)^2=\frac{2^{\alpha+\beta+1}\Gamma(\alpha+n+1)\Gamma(\beta+n+1)}{n!(\alpha+\beta+2n+1)\Gamma(\alpha+\beta+n+1)},
\label{IR2}
\end{eqnarray}
together with Eq.~\eqref{IR1}, yields the common contribution $T_E^1=T_O^1$ given in Eq.~\eqref{T^1}.

The second terms in Eqs.~\eqref{<H>n/2} and~\eqref{<H>n-1/2} involve integrals of the form
\begin{eqnarray}
\int_{-1}^{y_c}dy\,(1-y)^{\alpha}(1+y)^{\beta}\left(P_{n}^{\alpha,\beta}(y)\right)^2
&=&\int_{-1}^{0}dy\,(1-y)^{\alpha}(1+y)^{\beta}\left(P_{n}^{\alpha,\beta}(y)\right)^2\nb\\
&&+\int_{0}^{y_c}dy\,(1-y)^{\alpha}(1+y)^{\beta}\left(P_{n}^{\alpha,\beta}(y)\right)^2.
\label{T2I}
\end{eqnarray}
For the levels $n=0$ and $n=1$, the Jacobi polynomial has degree zero, and the required integral reduces to
\begin{eqnarray}
\int_{-1}^{y_c}dy\,(1-y)^{\alpha}(1+y)^{\beta}\left(P_{0}^{\alpha,\beta}(y)\right)^2
=\int_{-1}^{0}dy\,(1-y)^{\alpha}(1+y)^{\beta}+\int_{0}^{y_c}dy\,(1-y)^{\alpha}(1+y)^{\beta}.
\label{T2In0n1}
\end{eqnarray}
We use the following expression, which also supplies the terminal integral in the recursion for higher polynomial degrees:
\begin{eqnarray}
\int_{0}^{y_c}dy(1-y)^{\alpha+n}(1+y)^{\beta+n}\left(P_{0}^{\alpha+n,\beta+n}(y)\right)^2=\int_{0}^{y_c}dy(1-y)^{\alpha+n}(1+y)^{\beta+n}\nb\\
=2^{\alpha+\beta+2n+1}\left[\frac{\frac{1+y_c}{2}^{\beta+n+1}}{\beta+n+1}{_{2}F_1}\left(\beta+n+1,-\alpha-n,\beta+n+2,\frac{1+y_c}{2}\right)\right.\nb\\
\left.-\frac{\frac{1}{2}^{\beta+n+1}}{\beta+n+1}{_{2}F_1}\left(\beta+n+1,-\alpha-n,\beta+n+2,\frac{1}{2}\right)\right].\label{P_0int}
\end{eqnarray}
Setting the upper limit to $y_c=-1$ and changing the sign gives
\begin{eqnarray}
-\int_{0}^{-1}dy(1-y)^{\alpha}(1+y)^{\beta}\left(P_{0}^{\alpha,\beta}(y)\right)^2,
\end{eqnarray}
which is the first term on the right-hand side of Eq.~\eqref{T2In0n1}.
Combining the two intervals therefore yields
\begin{eqnarray}
\int_{-1}^{y_c}dy\,(1-y)^{\alpha}(1+y)^{\beta}\left(P_{0}^{\alpha,\beta}(y)\right)^2=\frac{2^{\alpha}(1+y_c)^{\beta+1}}{\beta+1}{_{2}F_1}\left(\beta+1,-\alpha,\beta+2,\frac{1+y_c}{2}\right).
\end{eqnarray}
For $n=0$, this becomes
\begin{equation}
\begin{aligned}
\int_{-1}^{y_c}dy\,(1-y)^{\lambda}(1+y)^{-\frac{1}{2}}=2^{\lambda+1}(1+y_c)^{\frac{1}{2}}{_{2}F_1}\left(\frac{1}{2},-\lambda,\frac{3}{2},\frac{1+y_c}{2}\right)
\end{aligned}
\label{rT2Ien0}
\end{equation}
and for $n=1$ it becomes
\begin{equation}
\begin{aligned}
\int_{-1}^{y_c}dy\,(1-y)^{\lambda-1}(1+y)^{\frac{1}{2}}=\frac{2^{\lambda}}{3}(1+y_c)^{\frac{3}{2}}{_{2}F_1}\left(\frac{3}{2},-\lambda+1,\frac{5}{2},\frac{1+y_c}{2}\right)
\end{aligned}
\label{rT2Ion1}
\end{equation}

For the remaining levels, the first integral on the right-hand side of Eq.~\eqref{T2I} can similarly be written as
\begin{eqnarray}
-\int_{0}^{-1}dy(1-y)^{\alpha}(1+y)^{\beta}\left(P_{n}^{\alpha,\beta}(y)\right)^2,
\label{T^21}
\end{eqnarray}
It is therefore sufficient to evaluate the second integral and then take its upper limit to $-1$ to recover the first.
Integration by parts gives
\begin{eqnarray}
\int_{0}^{y_c}dyf(y)g(y)=\left.g(y)\left(\int_{0}^{y}dkf(k)\right)\right|_{0}^{y_c}-\int_{0}^{y_c}dyg'(y)\left(\int_{0}^{y}dkf(k)\right),\label{intPart}
\end{eqnarray}
where
\begin{eqnarray}
f(y)=(1-y)^{\alpha}(1+y)^{\beta}P_{n}^{\alpha,\beta}(y),~~~g(y)=P_{n}^{\alpha,\beta}(y).
\end{eqnarray}
For the Jacobi polynomials, we use
\begin{eqnarray}
\int_{0}^{y}dk(1-k)^{\alpha}(1+k)^{\beta}P_{n}^{\alpha,\beta}(k)&=&\frac{1}{2n}\left[P_{n-1}^{\alpha+1,\beta+1}(0)-(1-y)^{\alpha+1}(1+y)^{\beta+1}P_{n-1}^{\alpha+1,\beta+1}(y)\right],\nb\\
\frac{d}{dy}P_{n}^{\alpha,\beta}(y)&=&\frac{\Gamma(\alpha+\beta+n+2)}{2\Gamma(\alpha+\beta+n+1)}P_{n-1}^{\alpha+1,\beta+1}(y).
\end{eqnarray}
Substitution into Eq.~\eqref{intPart} gives
\begin{eqnarray}
\frac{1}{2n}\left[P_{n}^{\alpha,\beta}(0)P_{n-1}^{\alpha+1,\beta+1}(0)-(1-y_c)^{\alpha+1}(1+y_c)^{\beta+1}P_{n}^{\alpha,\beta}(y_c)P_{n-1}^{\alpha+1,\beta+1}(y_c)\right]\nb\\
+\frac{\alpha+\beta+n+1}{4n}\int_{0}^{y_c}dy(1-y)^{\alpha+1}(1+y)^{\beta+1}\left(P_{n-1}^{\alpha+1,\beta+1}(y)\right)^2 .
\label{T3I}
\end{eqnarray}
For even parity, $n=2,4,\cdots$, this expression becomes
\begin{eqnarray}
\frac{1}{n}\left[P_{n/2}^{\lambda-n,-1/2}(0)P_{n/2-1}^{\lambda-n+1,1/2}(0)-(1-y_c)^{\lambda-n+1}(1+y_c)^{\frac{1}{2}}P_{n/2}^{\lambda-n,-1/2}(y_c)P_{n/2-1}^{\lambda-n+1,1/2}(y_c)\right]\nb\\
+\frac{\lambda-\frac{n}{2}+\frac{1}{2}}{2n}\int_{0}^{y_c}dy(1-y)^{\lambda-n+1}(1+y)^{\frac{1}{2}}\left(P_{n/2-1}^{\lambda-n+1,1/2}(y)\right)^2 .
\label{T2In/2}
\end{eqnarray}
For the remaining odd-parity levels, $n=3,5,\cdots$, it becomes
\begin{eqnarray}
\frac{1}{n-1}\left[P_{(n-1)/2}^{\lambda-n,1/2}(0)P_{(n-1)/2-1}^{\lambda-n+1,3/2}(0)-(1-y_c)^{\lambda-n+1}(1+y_c)^{\frac{3}{2}}P_{(n-1)/2}^{\lambda-n,1/2}(y_c)P_{(n-1)/2-1}^{\lambda-n+1,3/2}(y_c)\right]\nb\\
+\frac{\lambda-\frac{n}{2}+\frac{3}{2}}{2(n-1)}\int_{0}^{y_c}dy(1-y)^{\lambda-n+1}(1+y)^{\frac{3}{2}}\left(P_{(n-1)/2-1}^{\lambda-n+1,3/2}(y)\right)^2.
\label{T2In-1/2}
\end{eqnarray}
The remaining integral has the same form as the original one, with both exponents increased by one and the polynomial degree reduced by one.
Repeated substitution of Eq.~\eqref{intPart} into Eq.~\eqref{T3I} therefore terminates at degree zero and gives
\begin{eqnarray}
&&\frac{1}{2n}\left[P_{n}^{\alpha,\beta}(0)P_{n-1}^{\alpha+1,\beta+1}(0)-(1-y_c)^{\alpha+1}(1+y_c)^{\beta+1}P_{n}^{\alpha,\beta}(y_c)P_{n-1}^{\alpha+1,\beta+1}(y_c)\right]\nb\\
&&+\frac{\alpha+\beta+n+1}{4n}\frac{1}{2(n-1)}
\left[P_{n-1}^{\alpha+1,\beta+1}(0)P_{n-2}^{\alpha+2,\beta+2}(0)-(1-y_c)^{\alpha+2}(1+y_c)^{\beta+2}P_{n-1}^{\alpha+1,\beta+1}(y_c)P_{n-2}^{\alpha+2,\beta+2}(y_c)\right]\nb\\
&&+\frac{\alpha+\beta+n+1}{4n}\frac{\alpha+\beta+n}{4(n-1)}\left[\cdots\frac{\alpha+\beta+2}{4}\int_{0}^{y_c}dy(1-y)^{\alpha+n}(1+y)^{\beta+n}\left(P_{n-n}^{\alpha+n,\beta+n}(y)\right)^2\right],
\end{eqnarray}
The coefficient multiplying the final integral satisfies
\begin{eqnarray}
\frac{\alpha+\beta+n+1}{4n}\frac{\alpha+\beta+n}{4(n-1)}\frac{\alpha+\beta+n-1}{4(n-2)}\cdots\frac{\alpha+\beta+2}{4}=\frac{\Gamma(\alpha+\beta+n+2)}{4^n\Gamma(n+1)\Gamma(\alpha+\beta+2)},
\end{eqnarray}
and the final integral is given by Eq.~\eqref{P_0int}.
For even parity, Eq.~\eqref{T2In/2} consequently becomes
\begin{eqnarray}
&&\frac{1}{n}\left[P_{n/2}^{\lambda-n,-1/2}(0)P_{n/2-1}^{\lambda-n+1,1/2}(0)-(1-y_c)^{\lambda-n+1}(1+y_c)^{\frac12}P_{n/2}^{\lambda-n,-1/2}(y_c)P_{n/2-1}^{\lambda-n+1,1/2}(y_c)\right]\nb\\
&&+\frac{\lambda-\frac n2+\frac12}{2n(n-2)}
\left[P_{n/2-1}^{\lambda-n+1,1/2}(0)P_{n/2-2}^{\lambda-n+2,3/2}(0)-(1-y_c)^{\lambda-n+2}(1+y_c)^{\frac{3}{2}}P_{n/2-1}^{\lambda-n+1,1/2}(y_c)P_{n/2-2}^{\lambda-n+2,3/2}(y_c)\right]\nb\\
&&+\frac{\lambda-\frac n2+\frac12}{2n}\frac{\lambda-\frac n2-\frac12}{2(n-2)}\left\{\cdots\frac{\lambda-n+\frac32}{4}\frac{2^{\lambda+1-\frac{n}{2}}}{1+n}\left[{(1+y_c)}^{\frac{1}{2}+\frac{n}{2}}{_{2}F_1}\left(\frac{1}{2}+\frac{n}{2},-\lambda+\frac{n}{2},\frac{3}{2}+\frac{n}{2},\frac{1+y_c}{2}\right)\right.\right.\nb\\
&&\left.\left.-{_{2}F_1}\left(\frac{1}{2}+\frac{n}{2},-\lambda+\frac{n}{2},\frac{3}{2}+\frac{n}{2},\frac{1}{2}\right)\right]\right\},
\end{eqnarray}
while for odd parity Eq.~\eqref{T2In-1/2} becomes
\begin{eqnarray}
&&\frac{1}{n-1}\left[P_{(n-1)/2}^{\lambda-n,1/2}(0)P_{(n-1)/2-1}^{\lambda-n+1,3/2}(0)-(1-y_c)^{\lambda-n+1}(1+y_c)^{\frac32}P_{(n-1)/2}^{\lambda-n,1/2}(y_c)P_{(n-1)/2-1}^{\lambda-n+1,3/2}(y_c)\right]\nb\\
&&+\frac{\lambda-\frac n2+\frac{1}{2}}{2(n-1)(n-3)}
\left[P_{(n-1)/2-1}^{\lambda-n+1,3/2}(0)P_{(n-1)/2-2}^{\lambda-n+2,5/2}(0)-(1-y_c)^{\lambda-n+2}(1+y_c)^{\frac{5}{2}}P_{(n-1)/2-1}^{\lambda-n+1,3/2}(y_c)P_{(n-1)/2-2}^{\lambda-n+2,5/2}(y_c)\right]\nb\\
&&+\frac{\lambda-\frac n2+\frac{1}{2}}{2(n-1)}\frac{\lambda-\frac n2-\frac{1}{2}}{2(n-3)}\left\{\cdots\frac{\lambda-n+\frac52}{4}\frac{2^{\lambda+\frac{1}{2}-\frac{n}{2}}}{2+n}\left[(1+y_c)^{1+\frac{n}{2}}{_{2}F_1}\left(1+\frac{n}{2},-\lambda+\frac{1+n}{2},2+\frac{n}{2},\frac{1+y_c}{2}\right)\right.\right.\nb\\
&&\left.\left.-{_{2}F_1}\left(1+\frac{n}{2},-\lambda+\frac{1+n}{2},2+\frac{n}{2},\frac{1}{2}\right)\right]\right\}.
\end{eqnarray}
Taking $y_c=-1$ and changing the sign gives the contribution in Eq.~\eqref{T^21}.
For even parity, it is
\begin{eqnarray}
&&\frac{1}{n}\left(-P_{n/2}^{\lambda-n,-1/2}(0)P_{n/2-1}^{\lambda-n+1,1/2}(0)\right)+\frac{\lambda-\frac n2+\frac12}{2n(n-2)}
\left(-P_{n/2-1}^{\lambda-n+1,1/2}(0)P_{n/2-2}^{\lambda-n+2,3/2}(0)\right)\nb\\
&&+\frac{\lambda-\frac n2+\frac12}{2n}\frac{\lambda-\frac n2-\frac12}{2(n-2)}\left[\cdots\frac{\lambda-n+\frac32}{4}\frac{2^{\lambda+1-\frac{n}{2}}}{1+n}{_{2}F_1}\left(\frac{1}{2}+\frac{n}{2},-\lambda+\frac{n}{2},\frac{3}{2}+\frac{n}{2},\frac{1}{2}\right)\right] ,
\end{eqnarray}
and for odd parity it is
\begin{eqnarray}
&&\frac{1}{n-1}\left(-P_{(n-1)/2}^{\lambda-n,1/2}(0)P_{(n-1)/2-1}^{\lambda-n+1,3/2}(0)\right)+\frac{\lambda-\frac n2+\frac{1}{2}}{2(n-1)(n-3)}
\left(-P_{(n-1)/2-1}^{\lambda-n+1,3/2}(0)P_{(n-1)/2-2}^{\lambda-n+2,5/2}(0)\right)\nb\\
&&+\frac{\lambda-\frac n2+\frac{1}{2}}{2(n-1)}\frac{\lambda-\frac n2-\frac{1}{2}}{2(n-3)}\left[\cdots\frac{\lambda-n+\frac52}{4}\frac{2^{\lambda+\frac{1}{2}-\frac{n}{2}}}{2+n}{_{2}F_1}\left(1+\frac{n}{2},-\lambda+\frac{1+n}{2},2+\frac{n}{2},\frac{1}{2}\right)\right]
\end{eqnarray}
Combining the two integration intervals cancels the terms evaluated at zero.
The resulting integral for the remaining even-parity levels, $n=2,4,\cdots$, is
\begin{equation}
\begin{aligned}
&\int_{-1}^{y_c}dy\,(1-y)^{\lambda-n}(1+y)^{-\frac{1}{2}}\left(P_{n/2}^{\alpha,\beta}(y)\right)^2=\\
& -\frac{1}{n}
  (1-y_c)^{\lambda-n+1}(1+y_c)^{\frac12}
  P_{n/2}^{\lambda-n,-1/2}(y_c)
  P_{n/2-1}^{\lambda-n+1,1/2}(y_c) \\
& -\frac{\lambda-\frac n2+\frac12}{2n(n-2)}
  (1-y_c)^{\lambda-n+2}(1+y_c)^{\frac{3}{2}}
  P_{n/2-1}^{\lambda-n+1,1/2}(y_c)
  P_{n/2-2}^{\lambda-n+2,3/2}(y_c) \\
& -\cdots \\
& +\frac{\lambda-\frac n2+\frac12}{2n}
  \frac{\lambda-\frac n2-\frac12}{2(n-2)}
  \cdots
  \frac{\lambda-n+\frac32}{4}
  \frac{2^{\lambda+1-\frac{n}{2}}}{1+n}
  (1+y_c)^{\frac{1}{2}+\frac{n}{2}}
  {}_{2}F_1\!\left(
      \frac{1}{2}+\frac{n}{2},
      -\lambda+\frac{n}{2},
      \frac{3}{2}+\frac{n}{2},
      \frac{1+y_c}{2}
  \right) ,
\end{aligned}
\label{rT2Ie}
\end{equation}
For the remaining odd-parity levels, $n=3,5,\cdots$, the result is
\begin{equation}
\begin{aligned}
&\int_{-1}^{y_c}dy\,(1-y)^{\lambda-n}(1+y)^{\frac{1}{2}}\left(P_{(n-1)/2}^{\alpha,\beta}(y)\right)^2=\\
& -\frac{1}{n-1}
  (1-y_c)^{\lambda-n+1}(1+y_c)^{\frac32}
  P_{(n-1)/2}^{\lambda-n,1/2}(y_c)
  P_{(n-1)/2-1}^{\lambda-n+1,3/2}(y_c) \\
& -\frac{\lambda-\frac n2+\frac{1}{2}}{2(n-1)(n-3)}
  (1-y_c)^{\lambda-n+2}(1+y_c)^{\frac{5}{2}}
  P_{(n-1)/2-1}^{\lambda-n+1,3/2}(y_c)
  P_{(n-1)/2-2}^{\lambda-n+2,5/2}(y_c) \\
& -\cdots \\
& +\frac{\lambda-\frac n2+\frac{1}{2}}{2(n-1)}
  \frac{\lambda-\frac n2-\frac{1}{2}}{2(n-3)}
  \cdots
  \frac{\lambda-n+\frac52}{4}
  \frac{2^{\lambda+\frac{1}{2}-\frac{n}{2}}}{2+n}
  (1+y_c)^{1+\frac{n}{2}}
  {}_{2}F_1\!\left(
      1+\frac{n}{2},
      -\lambda+\frac{1+n}{2},
      2+\frac{n}{2},
      \frac{1+y_c}{2}
  \right) ,
\end{aligned}
\label{rT2Io}
\end{equation}

To proceed further, we consider discontinuities placed near the origin and near spatial infinity.
For first-order perturbation theory to describe the energy deformation accurately, the contributions in Eq.~\eqref{<H>} must remain small.
The first contribution is controlled by Eq.~\eqref{conPer}, while the second also depends on the position $y_c$ through Eqs.~\eqref{rT2Ie} and~\eqref{rT2Io}.
We consider the two limiting positions used in the main text.

Near the origin, $y_c\sim-1$, the Jacobi polynomials can be approximated by
\begin{eqnarray}
P_{n}^{\alpha,\beta}(y_c)\sim P_{n}^{\alpha,\beta}(-1)=(-1)^n\frac{\Gamma(\beta+n+1)}{\Gamma(n+1)\Gamma(\beta+1)},
\end{eqnarray}
This gives
\begin{equation}
\begin{aligned}
&(1-y_c)^{\alpha+m}(1+y_c)^{\beta+m}
P_{n-(m-1)}^{\alpha+m-1,\beta+m-1}(y_c)
P_{n-m}^{\alpha+m,\beta+m}(y_c) \\
&\quad \sim
(1-y_c^2)^m(1-y_c)^{\alpha}(1+y_c)^{\beta}
\frac{\Gamma(\beta+n+1)^2}
     {\Gamma(n-m+2)\Gamma(n-m+1)
      \Gamma(\beta+m)\Gamma(\beta+m+1)} \\
&\quad =
(1-y_c^2)^m(1-y_c)^{\alpha}(1+y_c)^{\beta}
\frac{1}{(n-m+1)(\beta+m)}
\left(
  \frac{\Gamma(\beta+n+1)}
       {\Gamma(n-m+1)\Gamma(\beta+m)}
\right)^2 \\
&\quad =
\frac{(1-y_c^2)^m}{(n-m+1)(\beta+m)}
\left(
  \frac{(n-m+1)_m}{(\beta)_m}
\right)^2
(1-y_c)^{\alpha}(1+y_c)^{\beta}
\left(
  \frac{\Gamma(\beta+n+1)}
       {\Gamma(n+1)\Gamma(\beta)}
\right)^2,
\end{aligned}
\end{equation}
where $(z)_m$ denotes the rising Pochhammer symbol.
To compare the contributions other than the last term in Eqs.~\eqref{rT2Ie} and~\eqref{rT2Io}, we evaluate the ratio of the $(m+1)$-th and $m$-th factors and obtain
\begin{eqnarray}
(1-y_c^2)\,\frac{(1+n-m)(n-m)}{(\beta+m)(\beta+m+1)} \, .\label{factorYC}
\end{eqnarray}
For both parity sectors, $\beta=\mp1/2$, the ratio is proportional to $1-y_c^2$.
Writing $y_c=-1+\epsilon$, the suppression is controlled by Eq.~\eqref{consty_c-1}.
In this regime, Eqs.~\eqref{rT2Ie} and~\eqref{rT2Io} can be approximated by retaining their first and last terms.

Near spatial infinity, $y_c\sim1$, we instead use
\begin{eqnarray}
P_{n}^{\alpha,\beta}(y_c)\sim P_{n}^{\alpha,\beta}(1)=\frac{\Gamma(\alpha+n+1)}{\Gamma(n+1)\Gamma(\alpha+1)},
\end{eqnarray}
which gives
\begin{eqnarray}
(1-y_c)^{\alpha+m}(1+y_c)^{\beta+m}P_{n-(m-1)}^{\alpha+m-1,\beta+m-1}(y_c)P_{n-m}^{\alpha+m,\beta+m}(y_c)\nb\\
\sim(1-y_c^2)^m(1-y_c)^{\alpha}(1+y_c)^{\beta}\frac{\Gamma(\alpha+n+1)\Gamma(\alpha+n+1)}{\Gamma(n-m+2)\Gamma(n-m+1)\Gamma(\alpha+m)\Gamma(\alpha+m+1)}\nb\\
=(1-y_c^2)^m(1-y_c)^{\alpha}(1+y_c)^{\beta}\frac{1}{(n-m+1)(\alpha+m)}\left(\frac{\Gamma(\alpha+n+1)}{\Gamma(n-m+1)\Gamma(\alpha+m)}\right)^2\nb\\
=\frac{(1-y_c^2)^m}{(n-m+1)(\alpha+m)}\left(\frac{(n-m+1)_{m}}{(\alpha)_{m}}\right)^2(1-y_c)^{\alpha}(1+y_c)^{\beta}\left(\frac{\Gamma(\alpha+n+1)}{\Gamma(n+1)\Gamma(\alpha)}\right)^2.
\end{eqnarray}
The corresponding ratio is
\begin{eqnarray}
(1-y_c^2)\,\frac{(1+n-m)(n-m)}{(\alpha+m)(\alpha+m+1)} \, .
\end{eqnarray}
For both parity sectors, $\alpha=\lambda-n$, this ratio is proportional to $1-y_c^2$ as $y_c\sim1$.
The same approximation, retaining the first and last terms of Eqs.~\eqref{rT2Ie} and~\eqref{rT2Io}, is then used.

Let us consider the large-index corrections for the fundamental-state.
For $n=0$, substituting Eqs.~\eqref{IR1},~\eqref{T^1}, and~\eqref{rT2Ien0} into Eq.~\eqref{<H>n/2} gives the fundamental-state correction in Eq.~\eqref{DeltaE0}.
For even $n\gg1$, only the first of the retained terms in Eq.~\eqref{rT2Ie} survives, since the coefficient of the last term vanishes in the limit
\begin{eqnarray}
\lim_{n\rightarrow{+\infty}}\frac{\Gamma(\alpha+\beta+n+2)}{4^n\Gamma(n+1)\Gamma(\alpha+\beta+2)}= 0 .
\end{eqnarray}
Using Eqs.~\eqref{IR1},~\eqref{T^1}, and~\eqref{<H>n/2}, we obtain
\begin{eqnarray}
\Delta E_n 
&=&(V_0-V_\mathrm{0+})\left[\frac{\lambda-n}{2\lambda+1}+4{(1-y_c^2)}(1-y_c)^{\lambda-n}(1+y_c)^{-\frac12}\left(\frac{\Gamma(\frac n2+\frac12)}{\Gamma(\frac n2+1)\Gamma(-\frac12)}\right)^2T_{\rm nc}\right]
\nb\\
&\simeq&(V_0-V_\mathrm{0+})\left[\frac{\lambda-n}{2\lambda+1}+2^{n-\lambda+\frac{3}{2}}\frac{\lambda-n}{n\pi}(1-y_c^2)(1-y_c)^{\lambda-n}(1+y_c)^{-\frac{1}{2}}\tan{\left(\pi\lambda-\frac{n\pi}{2}\right)}\right]
\label{ninfx0}
\end{eqnarray}
for $y_c\sim-1$, and
\begin{eqnarray}
\Delta E_n &=&(V_0-V_\mathrm{0+})\left[\frac{\lambda-n}{2\lambda+1}+\frac{(1-y_c^2)(1-y_c)^{\lambda-n}(1+y_c)^{-\frac12}}{2(\lambda-n+1)(\lambda-n)^2}\left(\frac{\Gamma(\lambda-\frac n2+1)}{\Gamma(\frac n2+1)\Gamma(\lambda-n)}\right)^2T_{\rm nc}\right]
\nb\\\nb\\
&\simeq&(V_0-V_\mathrm{0+})\left[\frac{\lambda-n}{2\lambda+1}+\frac{8^{n-\lambda-\frac{1}{2}}(\lambda-n)}{n(\lambda-n+1)\pi}(1-y_c^2)(1-y_c)^{\lambda-n}(1+y_c)^{-\frac{1}{2}}\tan{\left(\pi\lambda-\frac{n\pi}{2}\right)}\right]
\label{ninfxinf}
\end{eqnarray}
for $y_c\sim+1$, where
\begin{eqnarray}
T_{\rm nc}=2^{-\lambda+n+\frac 12}(\lambda-n)\frac{\Gamma(\frac n2+1)\Gamma(\lambda-\frac n2+\frac 12)}{\Gamma(\frac n2+\frac 12)\Gamma(\lambda-\frac n2+1)}.\nb
\end{eqnarray}
The reductions in Eqs.~\eqref{ninfx0} and~\eqref{ninfxinf} use the identities and large-$n$ approximations
\begin{eqnarray}
&&~~~~~~~~~\Gamma(z)\Gamma(1-z)=\frac{\pi}{\sin(\pi z)},~~~\Gamma(-\frac{1}{2})=-2\sqrt{\pi},~~~\Gamma(z)\Gamma(z+\frac{1}{2})=2^{1-2z}\sqrt{\pi}~\Gamma(2z),\nb\\
&&~~~~~~~~~~~~~~~~~~~~~~~~~~~~~~~~~~~~~~~~~~~~~\frac{\Gamma(\frac n2+\alpha)}{\Gamma(\frac n2+\beta)}\sim\left(\frac{n}{2}\right)^{\alpha-\beta},~~~~~~n\gg1\nb\\
&&\frac{\Gamma(\lambda-\frac n2+1)}{\Gamma(\frac n2+1)\Gamma(\lambda-n)}=\frac{\Gamma(n-\lambda+1)}{\Gamma(\frac n2+1)\Gamma(\frac{n}{2}-\lambda)}=\frac{(n-\lambda)\Gamma(n-\lambda)}{\Gamma(\frac n2+1)\Gamma(\frac{n}{2}-\lambda)}\simeq\frac{(n-\lambda)n^{\lambda}\Gamma(n-2\lambda)}{\Gamma(\frac n2+1)\Gamma(\frac{n}{2}-\lambda)}~~~~~~~~~~~~~~~~~~~~\nb\\
&&~~~~~~~~~=\frac{2^{n-2\lambda-1}(n-\lambda)n^{\lambda}}{\sqrt{\pi}}\frac{\Gamma(\frac{n}{2}-\lambda+\frac{1}{2})}{\Gamma(\frac{n}{2}+1)}\simeq\frac{2^{n-\lambda-\frac{1}{2}}(n-\lambda)}{\sqrt{n\pi}},~~~~~~n\gg1.
\end{eqnarray}
For a large index satisfying $n\gg\lambda$, the correction in Eq.~\eqref{ninfx0} is suppressed when
\bqn
(V_0-V_\mathrm{0+})\frac1n \ll 1 , 
\eqn
so that the perturbative treatment remains valid.
For Eq.~\eqref{ninfxinf}, writing $y_c=1-\epsilon$, the perturbation amplitude must satisfy the stronger condition
\bqn
(V_0-V_\mathrm{0+})\frac{\epsilon}{n^3}\left(\frac{8}{\epsilon}\right)^n \ll 1 ,
\eqn
for the correction to remain small compared with the unperturbed energy in Eq.~\eqref{EnPT}.

\section{The full expressions for the corrections to QNMs}\label{app2}

As elaborated in Sec.~\ref{section5}, if Eqs.~\eqref{rT2Ie} and~\eqref{rT2Io} are not approximated by retaining only the first and last terms, one derives the full expressions for the QNM corrections Eqs.~\eqref{Del_omey-1} and~\eqref{Del_omey+1} as follows.
For even parity, using Eqs.~\eqref{IR1},~\eqref{<H>n/2},~\eqref{T^1} and~\eqref{rT2Ie}, we have
\begin{equation}
\begin{aligned}
 &\Delta E_{n/2}=T_{E}^1+(V_0-V_{0+})\rm \frac{\frac{n}{2}!(\lambda-n)\Gamma\left(\lambda-\frac{n}{2}+\frac{1}{2}\right)}{2^{\lambda-n-\frac{1}{2}}\Gamma\left(\lambda-\frac{n}{2}+1\right)\Gamma\left(\frac{n}{2}+\frac{1}{2}\right)}\times Eq.~\eqref{rT2Ie} .
\end{aligned}
\label{<H>n/2full}
\end{equation}
In the case of $n=0$, one replaces Eq.~\eqref{rT2Ie} in the above expression by Eq.~\eqref{rT2Ien0}.
Similarly, for odd parity, using Eqs.~\eqref{IR1},~\eqref{<H>n-1/2},~\eqref{T^1} and~\eqref{rT2Io}, we find
\begin{equation}
\begin{aligned}
 &\Delta E_{(n-1)/2}=T_{O}^1+(V_0-V_{0+})\rm \frac{\frac{n-1}{2}!(\lambda-n)\Gamma\left(\lambda-\frac{n}{2}\right)}{2^{\lambda-n+\frac{1}{2}}\Gamma\left(\lambda-\frac{n}{2}+\frac{1}{2}\right)\Gamma\left(\frac{n}{2}+1\right)}\times Eq.~\eqref{rT2Io} .
\end{aligned}
\label{<H>n/2full}
\end{equation}
In the case of $n=1$, one replaces Eq.~\eqref{rT2Io} in the above expression by Eq.~\eqref{rT2Ion1}.
By further considering Eqs.~\eqref{delOme} and~\eqref{OmegaPT}, one obtains
\begin{equation}
\begin{aligned}
 &\Delta \omega_{E}=\frac{1}{2\Omega_E}\left[T_{E}^1+(V_0-V_{0+})\rm \frac{\frac{n}{2}!(\pi^{-1}(\lambda)-n)\Gamma\left(\pi^{-1}(\lambda)-\frac{n}{2}+\frac{1}{2}\right)}{2^{\pi^{-1}(\lambda)-n-\frac{1}{2}}\Gamma\left(\pi^{-1}(\lambda)-\frac{n}{2}+1\right)\Gamma\left(\frac{n}{2}+\frac{1}{2}\right)}\times Eq.~\eqref{rT2Ie}\right],
\end{aligned}
\label{Del_omey_full-1}
\end{equation}
\begin{equation}
\begin{aligned}
 &\Delta \omega_{O}=\frac{1}{2\Omega_O}\left[T_{O}^1+(V_0-V_{0+})\rm \frac{\frac{n-1}{2}!(\pi^{-1}(\lambda)-n)\Gamma\left(\pi^{-1}(\lambda)-\frac{n}{2}\right)}{2^{\pi^{-1}(\lambda)-n+\frac{1}{2}}\Gamma\left(\pi^{-1}(\lambda)-\frac{n}{2}+\frac{1}{2}\right)\Gamma\left(\frac{n}{2}+1\right)}\times Eq.~\eqref{rT2Io}\right].
\end{aligned}
\label{Del_omey_full+1}
\end{equation}

\bibliographystyle{h-physrev}
\bibliography{references_qian}

\end{document}